\long\def\symbolfootnote[#1]#2{\begingroup%
\def\thefootnote{\fnsymbol{footnote}}\footnote[#1]{#2}\endgroup} 
\def\aj{AJ}
\def\araa{ARA\&A}
\def\apj{ApJ}
\def\apjl{ApJ}
\def\apjs{ApJS}
\def\aap{A\&A}
\def\mnras{MNRAS}

\newcommand{\msun}{\textup{M}_{\odot}}
\newcommand{\mdot}{\dot{M}}
\newcommand{\msunyr}{\mbox{$M_{\odot}{\textup{yr}}^{-1}$}}
\newcommand{\mmin}{M_{\textup{\scriptsize{min}}}}
\newcommand{\mlim}{M_{\textup{\scriptsize{lim}}}}
\newcommand{\mmax}{M_{\textup{\scriptsize{max}}}}
\newcommand{\mmaxlt}{M_{\textup{\scriptsize{max}}}(\log\tau)}
\newcommand{\mup}{M_{\textup{\scriptsize{up}}}}

\newcommand{\tdis}{\mbox{$t_{\textup{\scriptsize{dis}}}$}}
\newcommand{\tdistot}{\mbox{$t_{\textup{\scriptsize{dis}}}^{\textup{\scriptsize{tot}}}$}}
\newcommand{\tdismc}{\mbox{$t_{*}^{\textup{\scriptsize{tot}}}$}}
\newcommand{\tdismi}{\mbox{$t_{j}^{\textup{\scriptsize{tot}}}$}}
\newcommand{\tdisfour}{\mbox{$t_{4}^{\textup{\scriptsize{tot}}}$}}

\newcommand{\trh}{t_{\textup{\scriptsize{rh}}}}

\newcommand{\dt}{\mbox{$\Delta t$}}
\newcommand{\rh}{r_{\textup{\scriptsize{h}}}}
\newcommand{\betafit}{\beta_{\textup{\scriptsize{fit}}}}

\newcommand{\dr}{\textup{d}}

\newcommand{\dndttext}{\dr N/\dr \tau}
\newcommand{\dndt}{\frac{\dr N}{\dr \tau}}
\newcommand{\dmdt}{\mbox{$\dot{M}$}}

\newcommand{\dndm}{\frac{\dr N}{\dr M}}
\newcommand{\dndmtext}{\dr N/\dr M}
\newcommand{\dndlogmtext}{\dr N/\dr\log M}
\newcommand{\dndlogttext}{\dr N/\dr\log\tau}
\newcommand{\dndmi}{\frac{\dr N}{\dr M_i}}
\newcommand{\dndmitext}{\dr N/\dr M_i}
\newcommand{\dmidm}{\frac{\partial M_i}{\partial M}}
\newcommand{\cfr}{\textup{CFR}}
\newcommand{\sfr}{\textup{SFR}}

\newcommand{\mvbright}{M_V^{\textup{\scriptsize{brightest}}}}
\newcommand{\mc}{M_*}

\newcommand{\lup}{L_{\textup{\scriptsize{up}}}}
\newcommand{\nup}{N_{\textup{\scriptsize{up}}}}
\newcommand{\mto}{M_{\textup{\scriptsize{TO}}}}
\newcommand{\cimfe}{\textup{CIMF}_{\textup{\scriptsize{emp}}}}
\newcommand{\muev}{\mu_{\textup{\scriptsize{ev}}}}
\newcommand{\mtot}{\Gamma\cdot\sfr\cdot\dt}

\documentclass[useAMS,usenatbib, times]{mn2e}
\usepackage{amssymb,latexsym,graphicx,natbib,eufrak,times}

\newcommand{\spaze}{\hspace{-0.05cm}}

\title[The star cluster mass function]
  {The early evolution of the star cluster mass function}
\author[M. Gieles]
  {M.~Gieles$^{1}$\footnote{}\\
  $^1$ European Southern Observatory, Casilla 19001, Santiago 19, Chile 
}
\date{Accepted 2009 January 6.  Received 2009 January 6; in original form 2008 December 4}

\def\LaTeX{L\kern-.36em\raise.3ex\hbox{a}\kern-.15em
    T\kern-.1667em\lower.7ex\hbox{E}\kern-.125emX}

\begin{document}         
\maketitle
\begin{abstract}
Several recent studies have shown that the star cluster initial mass function (CIMF) can be well approximated by a power law, with indications for  a steepening or truncation at high masses. This contribution considers the evolution  of such a mass function due to cluster disruption, with emphasis on the part of the mass function that is observable in the first $\sim$1\,Gyr. A Schechter type function is used for the CIMF, with a power law index of $-2$ at low masses and an exponential truncation at $\mc$. Cluster disruption due to the tidal field of the host galaxy and encounters  with giant molecular clouds flattens the low-mass end of the mass function, but there is always a part of the `evolved Schechter function'  that can be approximated by a power law with index $-2$. The  mass range for which this holds depends on age, $\tau$, and shifts to higher masses roughly as $\tau^{0.6}$.  
Mean cluster masses derived from luminosity limited samples increase with age  very similarly due to the evolutionary fading of clusters. Empirical mass functions  are, therefore, approximately power laws with index $-2$, or slightly steeper, at all ages.
The results are illustrated by an application to the star cluster population of the interacting galaxy M51, which can be well described by a model with $\mc=(1.9\pm0.5)\times10^5\,\msun$ and a  short (mass-dependent) disruption time destroying $\mc$ clusters  in roughly a Gyr.
\end{abstract}
\begin{keywords}
galaxies: star clusters --
open clusters and assocations: general --
globular clusters: general 
\end{keywords}

\section{Introduction}
The\blfootnote{$^\star$E-mail: mgieles@eso.org\\} survival chances of young ($\lesssim$1\,Gyr) star clusters and how this may, or may not, depend on their mass, $M$, and environment has been the topic of quite some debate recently. Two models for the early evolution of clusters have recently become available, with a very different role  of $M$ in the disruption description: a mass-dependent disruption  model (\citealt{2003MNRAS.338..717B}, hereafter BL03; \citealt{2005A&A...441..117L}; \citealt*{2005A&A...429..173L}) and a mass-independent disruption model \citep{2005ApJ...631L.133F, 2007AJ....133.1067W}. Both disruption models assume a continuous power law cluster initial mass function (CIMF) with index $-2$, based on empirical derivations of the mass function of young  ($\sim$10\,Myr) clusters with masses between $\sim$10$^2\,\msun$ and $\sim$10$^5\,\msun$ in a variety of galactic environments 
 (see for example \citet{1999ApJ...527L..81Z} for the case of the clusters in the Antennae galaxies; \citet{2003A&A...397..473B} for clusters in M51; \citet{2007ApJ...663..844M} for cluster in M82; \citet{2008MNRAS.383.1000D} for the SMC and \citet{2003MNRAS.343.1285D} for a compilation of clusters in different galaxies)\footnote{The observationally derived CIMF from young ($\lesssim$10\,Myr) clusters will be referred to as  the $\cimfe$. This is because the number of massive clusters younger than $\sim$10\,Myr is not necessarily sufficient to reveal the shape of the CIMF at high masses.  See Section~\ref{subsubsec:ev} for more details.}.

Based on a study of the age and mass distributions of star clusters in four different galaxies, BL03 conclude that the lifetime of star clusters, or disruption time-scale, $\tdis$, depends on $M$ as $\tdis\propto M^\gamma$, with $\gamma\simeq 0.62\pm0.06$ and with the proportionality constant dependent on galactic environment. This mass-dependent disruption (MDD) model is supported by our theoretical understanding  of the disruption of star clusters, since the same scaling of $\tdis$ with $M$, i.e. the same value of $\gamma$,  was found from $N$-body simulations of clusters dissolving under the combined effect of internal relaxation and external tides \citep{2003MNRAS.340..227B, 2008MNRAS.389L..28G}, and also for the disruption of clusters due to external perturbations, or `shocks'.  In fact, $\tdis$ due to shocks depends on both $M$ and  the half-mass radius, $\rh$, since $\tdis$ scales with the cluster density within  $\rh$ \citep{1958ApJ...127...17S, 1972ApJ...176L..51O}, which combined with the shallow dependence of  $\rh$ on $M$ for (young) clusters ($\rh\propto M^{0.1}$, e.g. \citealt{2004A&A...416..537L}), leads to a similar value of $\gamma$ as for tidal disruption.

MDD flattens the low-mass end of the cluster mass function, $\dndmtext$, at old ages. \citet{2001ApJ...561..751F} show that for a constant mass-loss rate ($\gamma=1$), the mass function of a single-age cluster population evolves towards a flat distribution at the low-mass end,  independent of the initial shape of the mass function.  Such a mass function is in good agreement with that of the Galactic globular cluster system. However, the globular cluster mass function (GCMF) is usually derived from the luminosity function using a constant mass-to-light  ratio, $M/L$. \citet{1991A&A...252...94M} find from dynamical mass estimates of Galactic globular clusters that  their $M/L$  is actually an increasing function of $M$, which could be the result of the preferential depletion of low-mass stars \citep{2008A&A...490..151K}. Such a mass-dependent $M/L$ alters the shape of the GCMF and could allow for smaller values of $\gamma$ (Kruijssen \& Portegies Zwart, in prep). A general expression for the logarithmic slope at low masses due to MDD is $\gamma-1$  \citep{2005A&A...441..117L}.
 
An MDD model for the early evolution of star clusters was introduced by  BL03. In the four galaxies studied by BL03, only masses for clusters in  M33 and M51 were available. The mass functions of clusters in these galaxies were presented for all ages, so no evidence for a flattening of the mass function for a selected set of old clusters was provided, although these integrated mass functions  are slightly flatter at low masses than a power law with index $-2$. \citet{2003ApJ...583L..17D} report a turnover in the mass function of clusters in region B of M82 and they claim that the population is approximately coeval ($\sim$1\,Gyr). This result was later  refuted, since from spectroscopic studies  a larger age range with a younger mean was found \citep{2008ApJ...674..846K}  and the turnover was attributed to detection incompleteness due to the  extent of clusters \citep{2007ApJ...667L.145S, 2008ApJ...679..404M}. \citet{2006MNRAS.366..295D} determine the mass function of LMC clusters in different age bins and find that it {\it steepens} with age, from a power law function with an index of $-1.8$ at $\sim$10\,Myr to one with an index of $-2.2$ at $\sim$1\,Gyr. \citet{2008A&A...487..557P}Ê also report a steeper mass function at older ages for Milky Way open clusters.  This steepening is contrary to what is expected from the evolution of a continuous power law CIMF with mass-dependent disruption. \citet{2004ApJ...613L.121G,2007AJ....133.2737G} show that the intermediate age (few Gyrs) cluster populations of the merger remnants NGC~1316 and NGC~3610, respectively, provide evidence for a turnover in their mass function, based on a flattening of the cluster luminosity function at low luminosities.  But these clusters have  already evolved over a significant fraction of a Hubble time, so,  it is not at all convincing that the mass function of clusters younger than a Gyr gets shallower than a power law with index $-2$.
 
\citet{1999ApJ...527L..81Z} determined the mass function of cluster in the Antennae  aged between 25 and 160\,Myr and show that it has essentially the same shape as that for the clusters between 2.5 and 6\,Myr, namely a power law  with index $-2$. \citet{2005ApJ...631L.133F} show that the age distribution, $\dndttext$, of mass limited sub-samples of clusters in the Antennae  galaxies declines as $\tau^{-1}$. Such an age distribution can result from  a constant formation history  combined with  a  90\% disruption  fraction  each age dex. Together with the similarity of the mass functions at different ages the authors concluded that the disruption of star clusters in the first Gyr is independent of their mass. Several claims have been made that this mass-independent disruption (MID) model also describes the age distribution of clusters in other galaxies, such as the SMC \citep{2006ApJ...650L.111C} and M33 \citep{2007AJ....134..447S}, but these results have later been ascribed to detection incompleteness, since luminosity limited cluster samples, not affected by disruption, also have an age distribution that scales approximately as  $\tau^{-1}$ \citep*{2007ApJ...668..268G}. 

The consequence of the 90\% MID model is that the number of clusters in logarithmic age bins is constant, which for a continuous power law CIMF results in a constant maximum cluster mass in such bins. \citet{2008A&A...482..165G}  showed that this is indeed the case for the Antennae, but  through a comparison to cluster populations in six other galaxies, they also show that the cluster population of the Antennae galaxies is unique in that sense. The MID model fails to reproduce the first few 100\,Myr of the (mass limited) $\dndttext$ in other galaxies (see the discussion in \citealt{2007ApJ...668..268G} and \citealt{2008A&A...482..165G}), so it seems not to be a `universal'  scenario for cluster evolution, as claimed by  \citet{2007AJ....133.1067W}.

In summary,  the MDD model can not reproduce the correct mass functions at old ages and the MID model can not reproduce the correct age distribution at young ages in galaxies other than the Antennae.

This study searches for the explanation of this problem by abandoning an assumption that both models make, namely a continuous  power law CIMF. Alternatively, a truncated \citet{1976ApJ...203..297S} function  (equation~\ref{eq:dndmi}) is considered  for the parent distribution from which cluster masses are drawn and this function  is evolved with mass-dependent disruption. In Section~\ref{sec:schechter} arguments for the choice of this truncated distribution function are given and its basic properties  are presented. In Section~\ref{sec:evschechter} an analytical model for evolved Schechter mass functions is presented and a comparison to the cluster population of M51 is provided in Section~\ref{sec:obs}. A discussion and conclusions are given in Section~\ref{sec:conclusions}.

\section{A Schechter function for the cluster initial mass function}
\label{sec:schechter}

\subsection{The need for a truncation}
\label{subsec:trun}
Assume that initial cluster masses, $M_i$,  are drawn from a \citet{1976ApJ...203..297S}  parent distribution function of the form

\begin{equation}
\dndmi=A\,M_i^{-2} \exp(-M_i/\mc),
\label{eq:dndmi}
\end{equation}
where $\mc$ is the mass where the exponential drop occurs and $A$ is a constant that scales with the cluster formation rate ($\cfr$). The constant $A$ can also be taken as a function of time, such that equation ~Ê(\ref{eq:dndmi}) describes the cluster formation history and thus represents the probability of a cluster with an initial mass between $M_i$ and $M_i+\dr M_i$ forming at a time between $t$ and $t+\dr t$.  Since $A$ will be assumed to be constant in most cases throughout this study and since the focus will be on the evolution of the mass function the notation $\dndmitext$ is preferred, though $\dr N/(\dr M_i\dr t)$ would be more precise.

As mentioned in footnote~1, this distribution function for the cluster initial mass function (CIMF) is not necessarily reflected from the empirically derived CIMF ($\cimfe$). This is because usually only  clusters younger than $\sim$10\,Myr are used to determine the $\cimfe$. This time interval is short compared to the range of cluster ages in a typical cluster population (few Gyrs). When sampling only a low number of clusters from the distribution function in equation~(\ref{eq:dndmi}), the most massive cluster actually formed, $\mmax$, can be less massive than $\mc$, and then the truncation is not recognised from the $\cimfe$. 
The analytical form of equation~(\ref{eq:dndmi}) was proposed for the luminosity function galaxies, where $L_*$ (instead of $\mc$) is the characteristic galaxy luminosity \citep{1976ApJ...203..297S}. This functional form follows from a stochastic self-similar model for the origin of galaxies from self-gravitating gas (e.g. \citealt{1974ApJ...187..425P}). On stellar scales a power law (self-similar) mass spectrum at low masses with an exponential cut-off at high masses also follows from theoretical models of fragmentation in a collapsing molecular  cloud \citep{1979ApJ...229..242S}.  
For a constant cluster formation efficiency the CIMF reflects the shape of the giant molecular clouds (GMCs) and cores from which the clusters form. Collisions between cores within the GMCs and destruction of the most massive cores by the star formation process also leads to a Schechter type (equilibrium) mass function for clusters \citep{1996ApJ...457..578M}.

The choice for a truncation is also motivated by several indications that the high-mass end of the cluster mass function is steeper when a larger age range, and thus a larger total number of clusters, is considered. The most important arguments are discussed below.

\subsubsection{The most massive cluster in logarithmic age bins}
\label{subsubsec:ev}

 \citet{2003AJ....126.1836H} studied the evolution of the most massive cluster mass, $\mmax$, in equally spaced logarithmic age bins, $\mmaxlt$, in the SMC and the LMC. They show that for a continuous power law mass function  $\mmaxlt$  increases with age.
This is because for a power law mass function with index $-\alpha$, $\mmax$ scales with the number of clusters, $N$, as $\mmax\propto N^{1/(\alpha-1)}$. Assuming a constant $\cfr$, the number of clusters per logarithmic unit of age, $\dndlogttext$, scales linearly with $\tau$, since $\dndlogttext\propto\tau\,\dndttext$, with $\dndttext$  constant. In this case $\mmaxlt\propto\tau^{1/(\alpha-1)}$.
 For $\alpha=2$ this results in a linear scaling of $\mmaxlt$ with $\tau$. 
\citet*{1993ApJ...404..144K} find such a relation in the $\log(M)$ vs. $\log(\tau)$ plane of clusters in the SMC, LMC, M31 and M33, which they attribute to time-dependent conditions for cluster formation.  \citet{1997ApJ...480..235E} were the first to suggest that such an increase is purely a statistical effect that arises from sampling  cluster masses from a power law function with index $-2$. This power law form was suggested for the mass function of cloud cores as the result of the fractal and turbulent nature of the interstellar gas \citep{1996ApJ...471..816E}. This scale-invariant structure of the gas combined with a near constant star formation efficiency then leads to the same simple power law form for the mass function of clusters  \citep{1997ApJ...480..235E}.
For a constant star formation rate ($\sfr$), the number of clusters is set by the time interval that is considered, which then determines the maximum cluster mass that can (on average) be expected. What this means is that the probability of finding the physical conditions, such as density and pressure, to form a massive cluster is determined by the structure of the gas, which can be described by a simple functional form.

 \citet{2003AJ....126.1836H}  find  $\mmaxlt\propto\tau^{0.7}$ in the SMC and LMC, from which they conclude that $\alpha\simeq2.4$.  However, from direct fits to the $\cimfe$ \citet{2006MNRAS.366..295D} and \citet{2008MNRAS.383.1000D} find indices of $\alpha=1.8\pm0.1$ and $\alpha=2.00\pm0.15$ for the LMC and the SMC, respectively.  The difference between the results of Hunter et al. and de Grijs and co-workers can be explained when  a Schechter distribution function (equation~\ref{eq:dndmi}) is assumed:  \citet{2003AJ....126.1836H} only use the most massive clusters, i.e. close to $\mc$ where the mass function is steeper, whereas the studies of de Grijs et al. use the full mass range of only the youngest clusters, for which $\mmax\lesssim\mc$. From this it can be concluded that in these galaxies it takes more than 10\,Myr  to sample enough clusters from the CIMF to reach $\mc$. This scenario is confirmed by the results of \citet{2006MNRAS.366..295D}  who show  that the mass function of LMC clusters  in different age bins gets steeper with age, from $-1.8\pm0.1$ at $\sim$10\,Myr to $-2.2\pm0.1$ at $\sim$1\,Gyr. Because these mass functions are constructed in equally spaced logarithmic age bins similar arguments hold as above: 
in the older age bins longer time intervals are considered, thus sampling higher masses from the CIMF where it is steeper. 
 Lastly, \citet{larsen08} recently showed that a Schechter function with $\mc=2\times10^5\,\msun$ provides a good description of the high-mass end of the mass function of LMC clusters when the full age range is considered (but excluding the globular clusters).
 
\citet{2008A&A...482..165G} also looked at $\mmaxlt$ for clusters in  the SMC and LMC (using the data from Hunter et al.) and added cluster populations from five other galaxies. A  linear increase of $\mmaxlt$ with $\tau$ holds for the Milky Way open clusters and for the clusters in the SMC, LMC, M33 and M83 up to ages of $\sim$100\,Myr, which implies that $\alpha=2$ for the mass range (sampled in that age range) in those galaxies. It also means that there has been no disruption of massive clusters in these galaxies in this age range. In the `universal' MID model of \citet{2007AJ....133.1067W}, 90\% of all clusters gets destroyed each age dex, which predicts a constant number of clusters in logarithmic age bins (if $\dndttext\propto\tau^{-1}$, then $\dr N/\dr \log\tau$=constant), and therefore a constant $\mmax$ in such bins \citep{2008A&A...482..165G}.
Since this is not the case in most of the galaxies, the linear scaling of $\mmaxlt$ with $\tau$ is an important argument against the `universal' MID model of \citet{2007AJ....133.1067W}.

When the full age range of the cluster population is considered, the increase of $\mmaxlt$ with $\tau$ is slower than linear, roughly $\mmaxlt\propto\tau^{0.7}$ \citep{2008A&A...482..165G},  in agreement with what Hunter et al. had found.
For clusters in M51 the increase of $\mmaxlt$ with $\tau$ is  even slower and for the Antennae galaxies  $\mmaxlt$ is essentially independent of $\tau$, suggesting that due to the high $\sfr$  of these galaxies $\mmax\simeq\mc$    already at short intervals of $\tau$, such that $\mmaxlt$ is approximately constant.  However,  a high MID fraction and a truncation in the CIMF have more or less the same effect on the evolution of $\mmaxlt$ with $\tau$ \citep{2008A&A...482..165G}. In addition, a non-constant formation history of clusters, one which was lower in the past, can also produce a constant $\mmaxlt$. Additional age and mass distributions are needed to tell these effects apart.

In Section~\ref{sec:obs}  it is shown that the truncated CIMF scenario combined with mass-dependent disruption  nicely explains the age distribution and mass function of M51 clusters and the MID model is rejected based on the age distribution of massive clusters.  However, a truncation in the mass function does not explain the $\tau^{-1}$ age distribution of the Antennae clusters.

\subsubsection{Variations in the cluster luminosity function}
\label{subsubsec:var}
Indirect evidence for a steepening of the cluster mass function at the high-mass end comes from the luminosity function (LF) of clusters.   The LFs of clusters in different galaxies can be well approximated by  a power law, but with an index smaller  than the index of the $\cimfe$ (between $-2.5$ and $-2$, e.g. \citealt{2002AJ....123..207D, 2002AJ....123.1381E, 2002AJ....124.1393L}) with the LF being steeper at higher luminosities \citep{1999AJ....118.1551W, 1999AJ....118..752Z, 2002AJ....123.1411B, 2002AJ....124.1393L, 2005A&A...443...41M, 2006A&A...450..129G, 2006A&A...446L...9G, 2008AJ....135.1567H}. The LF consists of clusters with different ages. Due to the age dependent light-to-mass ratio of clusters, the LF does not necessarily have the same shape as the mass function. However, {\it if} the CIMF is a continuous power law with the same index at all ages,  i.e. with the physical maximum much higher than $\mmax$, the LF will be a power law with the exact same index. Age dependent extinction or bursts in the formation rate would not cause a difference between the CIMF and the LF. An addition of identical power laws always results in the same power law. So the fact that the LF is slightly steeper than the $\cimfe$ is already a strong indication that the CIMF is not a continuous power law function.

When an abrupt truncation of the CIMF at some mass $\mup$ is assumed, it is  possible to roughly estimate  the index of the bright-end of the LF. Assume that the CIMF is fully populated, i.e. the mass of the most massive cluster actually formed, $\mmax$, is equal to $\mup$.
 Then  assume a constant formation history of clusters, so  a constant number of the most massive clusters per unit of time: $\dr\nup/\dr \tau=$constant. The luminosities of these clusters, $\lup$, however, depend strongly on age. The light-to-mass ratio, or the flux of a cluster of constant mass, scales roughly with $\tau$ as $\tau^{-\zeta}$,  with $0.7\lesssim\zeta\lesssim1$ depending on the filter, 
 such that $\partial\lup/\partial\tau\propto\tau^{-\zeta-1}\propto\lup^{1+1/\zeta}$.
Then the  LF of such clusters is
\begin{eqnarray}
\frac{\dr \nup}{\dr \lup}& \propto &\frac{\dr \nup}{\dr\tau}\left|\frac{\partial\,\tau}{\partial \lup} \right|\\
	                    & \propto & \lup^{-1-1/\zeta},\\
	                    & \propto & \lup^{-2.5},	                    
 \label{eq:dndlindex}
\end{eqnarray}
where in the last step $\zeta=0.7$ is used  (BL03, \citealt{2007ApJ...668..268G}), such that the index of $-2.5$ holds for the $V$-band LF.   The same arguments hold for the luminosities of the 2$^\textup{\scriptsize{nd}}$, 3$^\textup{\scriptsize{d}}$, etc. most massive clusters, such that the bright-end of the LF of the entire population is an addition of power law with index $-2.5$, resulting in a power law with index $-2.5$ .

The faint-end of the LF should still be  a power law with index $-2$ or flatter if MDD is important. This double power law shape for the LF  was found by \citet{2006A&A...446L...9G} and \citet{1999AJ....118.1551W} for the LF of clusters in M51 and the Antennae, respectively. When a Schechter function for the CIMF is considered, the logarithmic slope of the LF shows a smooth decline from $-2$ to roughly $-3$ between $M_V\simeq-4$ and $M_V\simeq-12$   \citep{larsen08}. MDD affects the faint-end ($M_V\gtrsim-8$) of the LF and makes it slightly shallower (index $>-2$).

\subsubsection{The value of $\mc$}
Is the value of $\mc$ universal? Or does it somehow depend on galactic conditions, such as the $\sfr$?
The two proxies for the shape of the high-mass end of the CIMF, namely the evolution of $\mmaxlt$  with $\tau$  (Section~\ref{subsubsec:ev}) and the power law index of the LF (Section~\ref{subsubsec:var}), both show more convincing signatures of a truncation in the mass function in galaxies with a high SFR. This suggests that in such environments there are numerous clusters with masses above $\mc$, as opposed to galaxies with a low SFR. This excludes a linear dependence of $\mc$  on the $\sfr$.
 If $\mc$ would scale linearly with the $\sfr$, then it would take the same amount of time in each galaxy for $\mmax$ to reach $\mc$, since in the power law regime of the Schechter function $\mmaxlt\propto\tau$ (Section~\ref{subsubsec:ev}).  This was suggested by \citet*{2004MNRAS.350.1503W} and \citet{2007MNRAS.379...34M}, who claim a formation epoch of $10\,$Myr for an entire CIMF. If the CIMF is populated up to the highest mass each 10\,Myr, than there should be no increase of $\mmaxlt$ with age (assuming all age bins have a width larger than 10\,Myr). But this is not what  \citet{2003AJ....126.1836H} and \citet{2008A&A...482..165G} find.   \citet{2004MNRAS.350.1503W} base their result on the relation between the luminosity of the  brightest cluster in a galaxy, $\mvbright$, with the $\sfr$ of that galaxy using data from \citet{2002AJ....124.1393L} and the assumption that the brightest cluster is also the most massive cluster. They then use a constant light-to-mass ratio to convert luminosities to masses.  This assumption was further investigated by \citet{larsen08}, who determined the ages of all $\mvbright$ clusters from \citet{2002AJ....124.1393L} and showed that the brightest clusters  ($\mvbright\lesssim-11$) are young ($\lesssim$30\,Myr) while the clusters with $\mvbright\gtrsim-11$ have a large age spread (between $\sim$10\,Myr and $\sim$1\,Gyr). If the age of $\mvbright$ is $M_V$ dependent, then so is the light-to-mass ratio. Assuming that it is constant can thus lead to a misinterpretation of the data.  \citet{larsen08} showed that the $\mvbright$ vs. $\sfr$ relation for (quiescent) spiral galaxies is consistent with a constant $\mc$ of $\mc\simeq2\times10^5\,\msun$.
 
This value of $\mc$ is probably not universal,  since there are extremely massive clusters known ($10^7-10^8\,\msun$, \citealt{2004A&A...416..467M,2006A&A...448..881B}), which  are unlikely to have  formed from a Schechter function with $\mc=2\times10^5\,\msun$.  Jord{\'a}n~et~al.~(\citeyear{2007ApJS..171..101J}, hereafter J07) determine $\mc$ values of globular cluster populations of early-type galaxies in the Virgo Cluster by comparing the cluster luminosity functions to `evolved Schechter functions'. They find a trend of $\mc$ with host galaxy luminosity, with $\mc$ being higher in brighter galaxies, and values for $\mc$ between a few times $10^5\,\msun$ and a few times $10^6\,\msun$.  Several other studies have also shown that  a Schechter function fits the high-mass end of the globular cluster mass function  better  than a continuous power law with index $-2$ (\citealt{1996ApJ...457..578M,2000ApJ...542L..95B, 2001ApJ...561..751F, 2006ApJ...650..885W, 2008arXiv0811.1437H}).

\citet{{2002AJ....123.1454B}} suggest there should exist a maximum cluster mass based on the \citet{1998ApJ...498..541K} relation between the $\sfr$ and gas density and the assumption of a pressure equilibrium between  the ambient interstellar medium and the cloud-cores from which star clusters form. For a constant cluster density this relation is $\mup\propto\sfr^2$ and for a constant cluster radius, it would be $\mup\propto\sfr^{2/3}$. With the latter increase of $\mup$ with the $\sfr$, i.e. slower than linear, it is possible to explain the observational trends  discussed in Sections~\ref{subsubsec:ev}--\ref{subsubsec:var}. The key point is that $\mc$ is probably environment dependent and for the masses of a cluster population in a given galaxy to be affected by the truncation, the product of the $\sfr$ times the age range has to be high. For $\mc\propto\sfr^{2/3}$, low $\sfr$ galaxies need longer time intervals for $\mmax$ to reach $\mc$ than high $\sfr$ galaxies, which seems to be what is found from observations. 
The relation between $\mmax$ and the $\sfr$ will be quantified in more detail in Section~\ref{subsec:prop}.

\subsection{The relation between $\mmax$ and the star formation rate}
\label{subsec:prop}

 \begin{figure}
 \includegraphics[width=9cm]{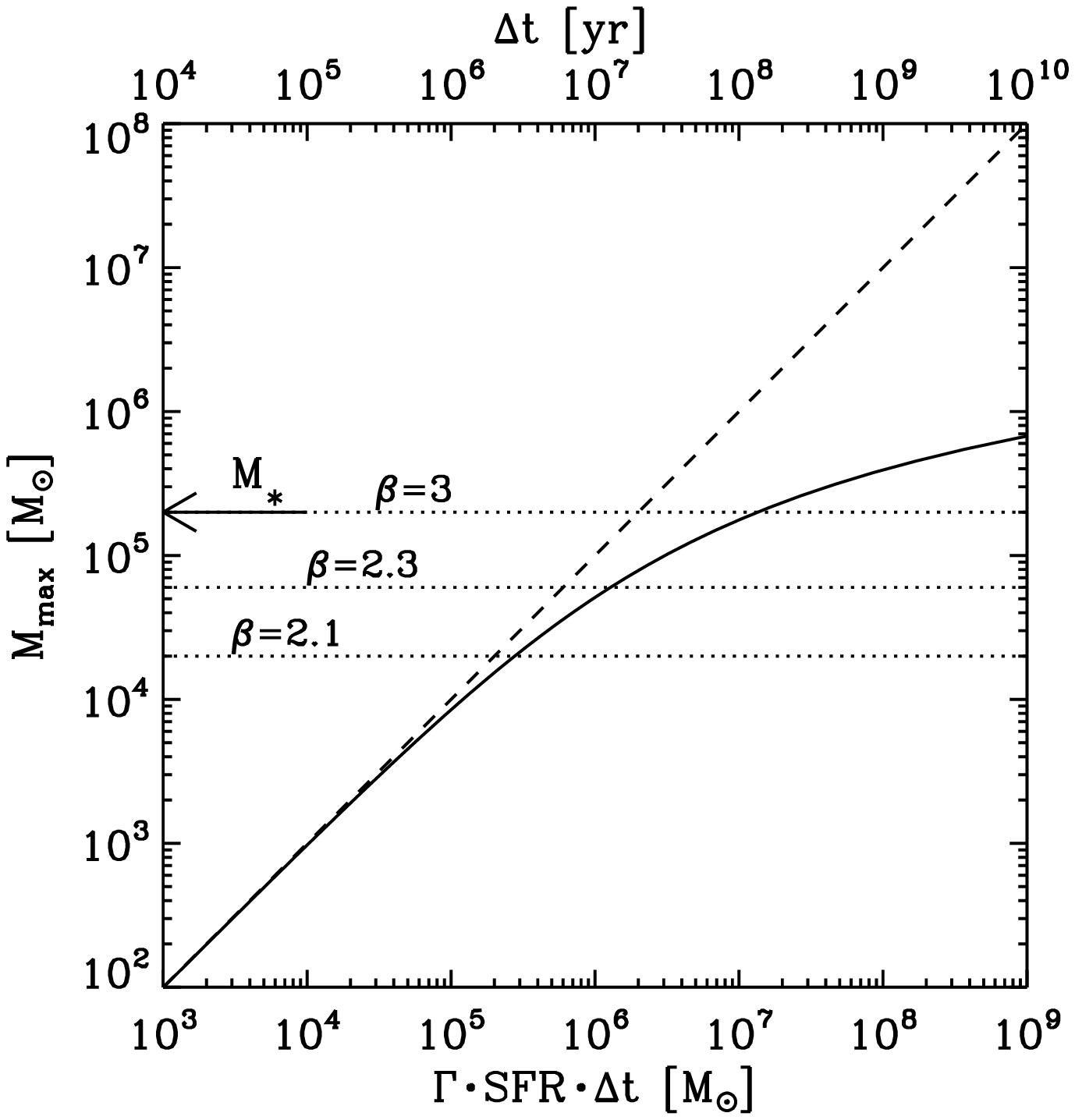}  	
     \caption{Example of the relation between the mass of the most massive cluster, $\mmax$,  as a function of the total mass formed in clusters, $\mtot$, when the CIMF is a Schechter function (equation~\ref{eq:dndmi}) with $M_*=2\times10^5\,\msun$ \citep{larsen08}. The dashed line shows a prediction for $\mmax$ in the case of an nontruncated mass function. The dotted lines indicate three masses and the corresponding logarithmic slope, $-\beta$, of the Schechter function. The top $x$-axis shows the time range that is needed to form the amount of mass that is represented on the bottom $x$-axis when $\sfr=1\,\msunyr$ and $\Gamma=0.1$.}
   \label{fig:ex}
\end{figure}

The mass of the most massive cluster in a cluster population, $\mmax$, depends on how many cluster are formed. This can be related to the $\cfr$ which scales linearly with the constant $A$  in equation~(\ref{eq:dndmi}), since

\begin{eqnarray}
\cfr&=&\int_{\mmin}^\infty M_i\,\dndmi\dr M_i,\\
	&=&A\,{\textup E}_{1}\left(\frac{\mmin}{\mc}\right), \label{eq:mtot00}\\
	&\simeq&10\,A,
	\label{eq:mtot0}
\end{eqnarray}
where ${\textup{E}}_n(x)$ with $n=1$ is a generalised expression of the exponential integral\footnote{The exponential integral is defined as ${\textup E}_n=\int_{1}^\infty t^{-n}\exp(-xt)\dr t$, and is related to the (upper) incomplete Gamma function, $\Gamma(a,x)$, as ${\textup E}_n(x)=x^{n-1}\,\Gamma(1-n,x)$. See Section~6.3 in \texttt{Numerical Recipes} \citep{1992nrfa.book.....P} for details on the exponential integral and how to implement this function in a code. The exponential integral is predefined in \texttt{IDL} as the function \texttt{expint(n,x)}.}. In the last step  $\mmin/\mc\simeq10^{-5}$ is used. For a ratio of $\mmin/\mc=10^{-4}(10^{-3})$ a relation of  $\cfr\simeq8A(6A)$ is found, showing that the relation between the $\cfr$ and $A$ is relatively insensitive to the ratio $\mmin/\mc$. The variable $A$ can be a function of time, $A(t)$, to include variations in the $\cfr$. This does not influence the shape of the mass function, but it will affect the shape of the age distribution, which can be derived once the CIMF is evolved with disruption (Section~\ref{sec:evschechter}). 

The relation between $\mmax$ and $A$ can be found from
\begin{eqnarray}
1&=&\int_{\mmax}^\infty \dndmi\,\dr M_i,\\
 &=&\frac{A\,{\textup E}_2(\mmax/\mc)}{\mmax}.
	\label{eq:mmax}
\end{eqnarray}
For $\mmax<<\mc$ the term ${\textup E}_2(\mmax/\mc)\simeq1$, and $A\simeq\mmax$, which is the result for a continuous power law with index $-2$.  An expression for $\mmax$ as a function of the $\cfr$ can be found by substituting $A$ as a function of the $\cfr$ from equation~(\ref{eq:mtot00}) in equation~(\ref{eq:mmax}). Since this is then in units of $\msunyr$, a multiplication  by some time interval, $\dt$, is necessary  to get an expression for $\mmax$ in $\msun$. Assuming that a constant fraction $\Gamma$ of the $\sfr$  ends up in star clusters that survive the embedded phase \citep{bastian08}, i.e. $\cfr=\Gamma\cdot\sfr$, then a relation between $\mmax$ and the $\sfr$ is found

\begin{eqnarray}
\frac{\mmax}{{\textup E}_2(\mmax/\mc)}&=&\frac{\Gamma\cdot\sfr\cdot\dt}{{\textup E}_{1}\left(\mmin/\mc\right)}.
\label{eq:mmaxmtot}
\end{eqnarray}

This relation between $\mmax$ and $\mtot$ is illustrated  in Fig.~\ref{fig:ex}. Since the exponential integral is not easily inverted  the product $\mtot$ is calculated for a series of $\mmax$ values. A value of $\mc=2\times10^5\,\msun$ is used, which was found for spiral galaxies by \citet{larsen08}, and ${\textup E}_1(\mmin/\mc)=10$, corresponding to $\mmin\simeq1$. The dotted lines indicate three masses where the logarithmic slope of the mass function, $-\beta$,  has the values $-2.1, -2.3$ and $-3$.
The logarithmic slope at $M$ of a mass function $\dndmtext$ is defined as
\begin{eqnarray}
-\beta(M)&\equiv&\frac{\dr\ln(\dndmtext)}{\dr\ln M},
\label{eq:betadef}
\end{eqnarray}
which for the CIMF gives
\begin{equation}
-\beta(M_i)=-2-M_i/\mc.
\label{eq:betadef2}
\end{equation}
Note that  $\beta$ is defined  such that it is positive for a declining $\dndmtext$.

 \begin{figure*}
 \includegraphics[width=8.cm]{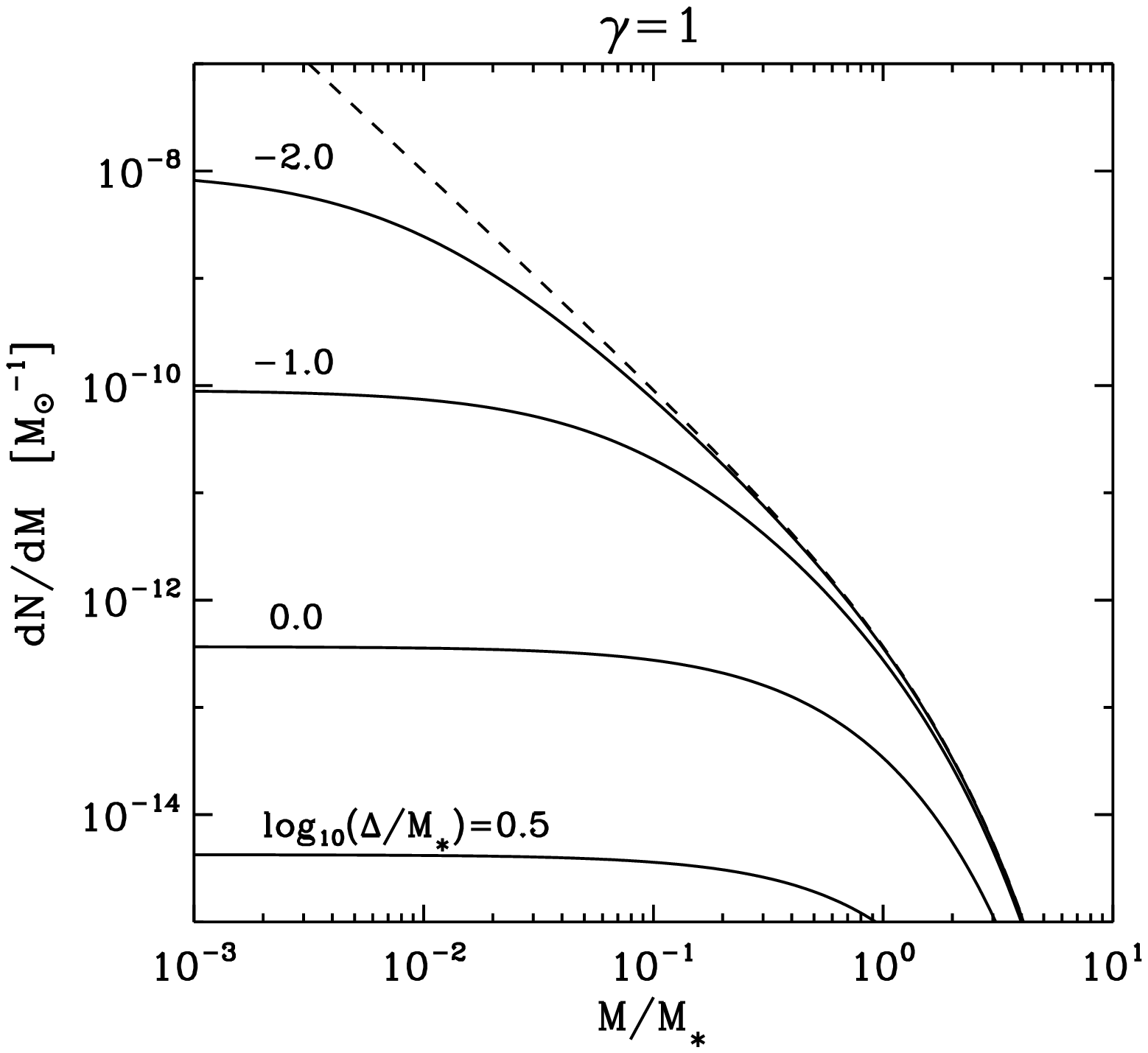} \includegraphics[width=8.cm]{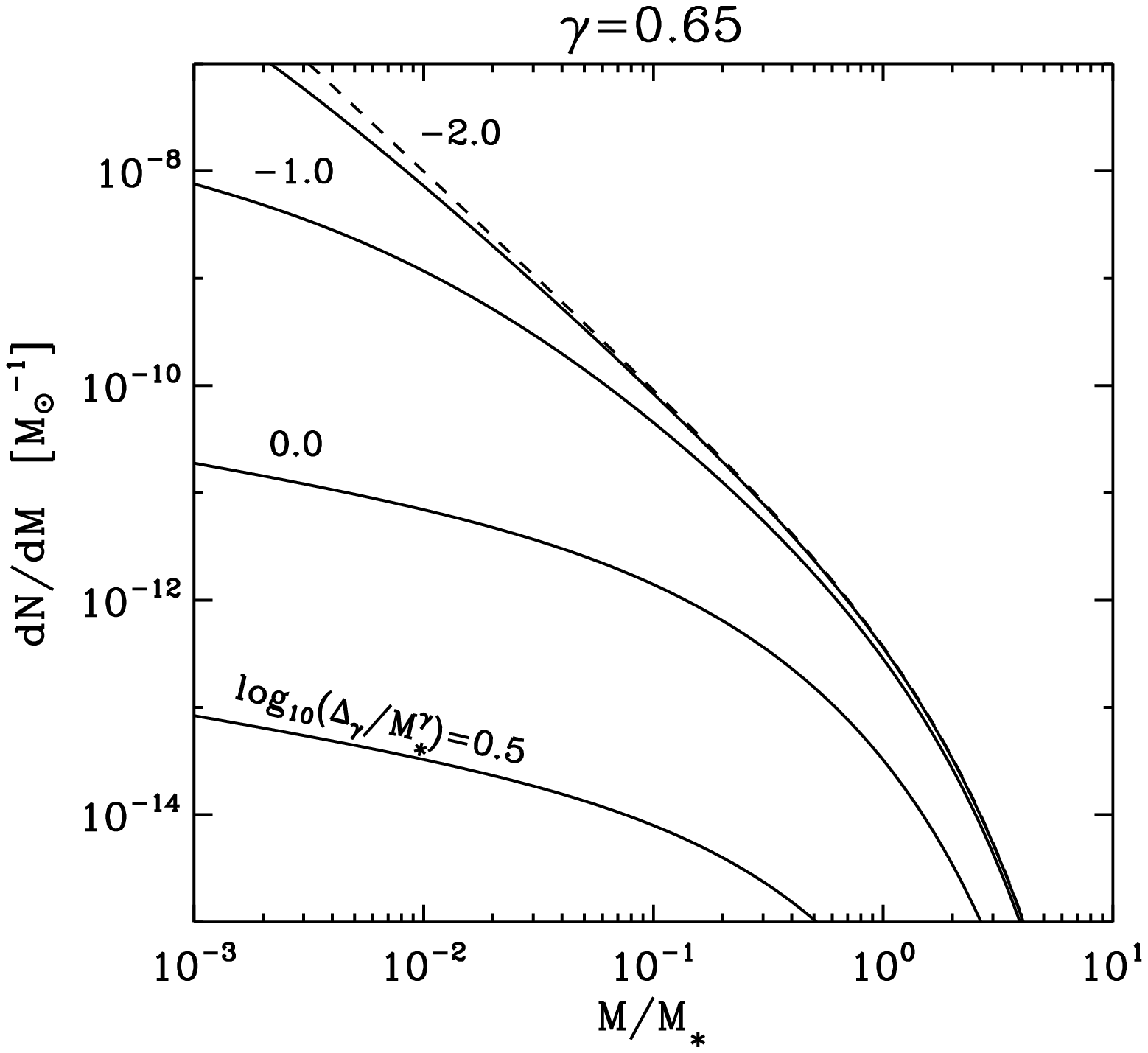}	
     \caption{Different stages of the evolved Schechter mass function for a constant mass loss rate (left panel, equation~[\ref{eq:dndmjordan}]) and a mass-dependent mass loss rate (right panel, equation~Ê[\ref{eq:dndmnonlin}]). The corresponding values of the ratio $\Delta/\mc$ (left) and $\Delta_\gamma/\mc^\gamma$ (right) are indicated. The dashed line in both panels shows the Schechter function at $t=0$ (equation~\ref{eq:dndmi}). }
   \label{fig:dndm}
\end{figure*}

From Fig.~\ref{fig:ex} we see that  $\mmax\simeq0.1\,\mtot$ for $\mmax\lesssim0.1\,\mc$. This corresponds to the mass range of the CIMF where $\beta\simeq2$. For higher values of $\mtot$ the relation between $\mmax$ and $\mtot$ flattens. The dashed line shows the predicted $\mmax$ for a nontruncated mass function, or a much higher $\mc$. Even though the two predictions diverge to a difference up to two dex in $\mmax$, a large increase of $\mtot$ (three dex) is needed to reach this difference. When $\mmax$ in the $\cimfe$ is sampled close to or a bit above $\mc$, the difference between a continuous power law and a Schechter function will be hard to tell. Mainly because the number of clusters in this high-mass tail is  low.

The top $x$-axis of Fig.~\ref{fig:ex} is labelled with $\dt$ values corresponding to $\mtot$ values with a fixed $\sfr$ and $\Gamma$ of $\sfr=1\,\msun\,$yr$^{-1}$ and $\Gamma=0.1$. These $\dt$ values can be interpreted as age ranges. In this illustrative example of a galaxy with a moderate $\sfr$, the $\cimfe$ ($\dt\lesssim10\,$Myr) will contain clusters with masses up to $\sim$5$\times10^4\,\msun$, while clusters with masses up to $\sim$10$^6\,\msun$ have formed when an age range of $\sim$10\,Gyr is considered. However, without the exponential truncation clusters with masses up to $10^8\,\msun$ are predicted to have formed. The example in Fig.~\ref{fig:ex} can be applied to the Milky Way. Its most massive  young star cluster known to date  is Westerlund~1 which has a mass around $10^5\,\msun$ \citep{2005A&A...434..949C, 2007A&A...466..151M}.
There are a handful of other massive clusters in the Milky Way known, such as the Arches and Quintuplet cluster towards the Galactic centre, NGC~3603 and the two recently discovered red super giant clusters \citep{2006ApJ...643.1166F, 2007ApJ...671..781D}. They  are all around $\sim$$10^4\,\msun$ and span an age range of $\sim$$10\,$Myr. From Fig.~\ref{fig:ex} it can be seen that for this age range values of $1-5\times10^4\,\msun$ are expected for $\mmax$. If the last Gyr of cluster formation is considered, the Galaxy has probably formed several clusters more massive than Westerlund~1, but not as many, nor as massive compared to the scenario in which the CIMF is a continuous power law.

With the properties of the CIMF introduced, an analytical expression for the `evolved Schechter function', i.e. the cluster mass function after mass-dependent disruption is applied, can now be derived.

\section{The evolved Schechter function}
\label{sec:evschechter}

In this section the `evolved Schechter function'  is presented, based on the CIMF of equation~(\ref{eq:dndmi}) which is evolved with mass-dependent cluster disruption  due to e.g. the tidal field of the host galaxy and/or encounters with GMCs. Cluster mass loss due to stellar evolution is not considered  at this stage. Also, the power law part of the mass function is always assumed to have an index of $-2$  for the moment.

It is a simple mathematical exercise to derive all expressions in this section for a variable index  and to include the effect of mass loss by stellar evolution, but for reasons of simplicity this is not done in the derivations of the formulae in the coming two sections. In Section~\ref{subsec:full} these two effects are added to the formalism.

\subsection{Constant mass loss rate}
\label{subsec:lin}
The evolved Schechter function is first introduced in  the form described by J07. The mass loss rate of clusters, $\dmdt$, is assumed to be constant. The disruption time for a cluster of mass $M$, i.e. the time needed  to reach $M=0$, is then given by 

\begin{eqnarray}
\tdis&\equiv&\frac{M}{\dmdt},\\
     &=&t_0\,M, 
\label{eq:tdislin}
\end{eqnarray}
where in the second step  $t_0\equiv 1/\dmdt$ is introduced. This  linear scaling of $\tdis$ with $M$ follows from the assumption that $\tdis$ is a constant times the half-mass relaxation time, $\trh$, and the assumption of a constant  cluster half-mass density  and a constant Coulomb logarithm in $\trh$  (e.g.~\citealt{1997ApJ...474..223G, 2001ApJ...561..751F}). A more general expression for the scaling of $\tdis$ with $M$ of equation~(\ref{eq:tdislin}) would be $\tdis=t_0M^\gamma$ (equation~\ref{eq:tdis}), with $\gamma=1$. 

The mass evolution of a cluster with time\footnote{To keep the formulae synoptic, the variables $M$ and $\dndmtext$, i.e. without the subscript $i$, are used for to the mass and the mass function as a function of time, i.e. $M(t)$ and $\dndmtext(t)$, respectively. The variables $M$ and $M_i$ are in units of $\msun$ and $t$ and $t_0$ are in units of time (Myr).}
 is then
\begin{eqnarray}
M&=&M_i-\dot{M}t,\nonumber\\
       &=&M_i-\Delta,
\label{eq:mtlin}
\end{eqnarray}
where  the variable $\Delta\equiv t/t_0 (=\dmdt t)$ is introduced, which is a proxy of time, but has the dimension of mass since it is the amount of mass with which  $M_i$ has reduced at time $t$ due  to disruption.  
 
Though the assumptions that have to be made to arrive to this constant mass loss are arguable, the result is a mathematically appealing description of the evolution of cluster masses, enabling us to make simple analytical predictions. In Section~\ref{subsec:nonlin} a more realistic $\dmdt$, based on results of $N$-body simulations, is explored.

The mass function as a function of  time, $\dndmtext$, follows from the conservation of number
\begin{equation}
\dndm = \dndmi\left|\dmidm\right|.
\label{eq:dndmdef}
\end{equation}
Since  $\partial M_i/\partial M=1$ and $M_i=M+\Delta$ (equation~\ref{eq:mtlin}) it then follows that

\begin{equation}
\dndm = \frac{A}{\left[M+\Delta\right]^{2}}\exp\left(-\frac{M+\Delta}{\mc}\right).
\label{eq:dndmjordan}
\end{equation}
The behaviour of this evolved Schechter function is shown in the left panel of Fig.~\ref{fig:dndm}. The dashed line shows the CIMF (equation~[\ref{eq:dndmi}] or equation~[\ref{eq:dndmjordan}] with  $\Delta=0$) and the full lines show the result of equation~(\ref{eq:dndmjordan}) at different times (i.e. for various ratios of $\Delta/\mc$). For $M\lesssim 0.1\Delta$  and $\Delta\lesssim\mc$ the $\dndmtext$ is flat ($\beta=0$). This is the direct consequence  of the constant $\mdot$ \citep{2001ApJ...561..751F}. The reader is referred to J07 for an equivalent functional form describing the luminosity functions, i.e. the number of clusters in constant magnitude interval, which is peaked and rises at low luminosities.

\begin{figure*}
 \includegraphics[width=8.cm]{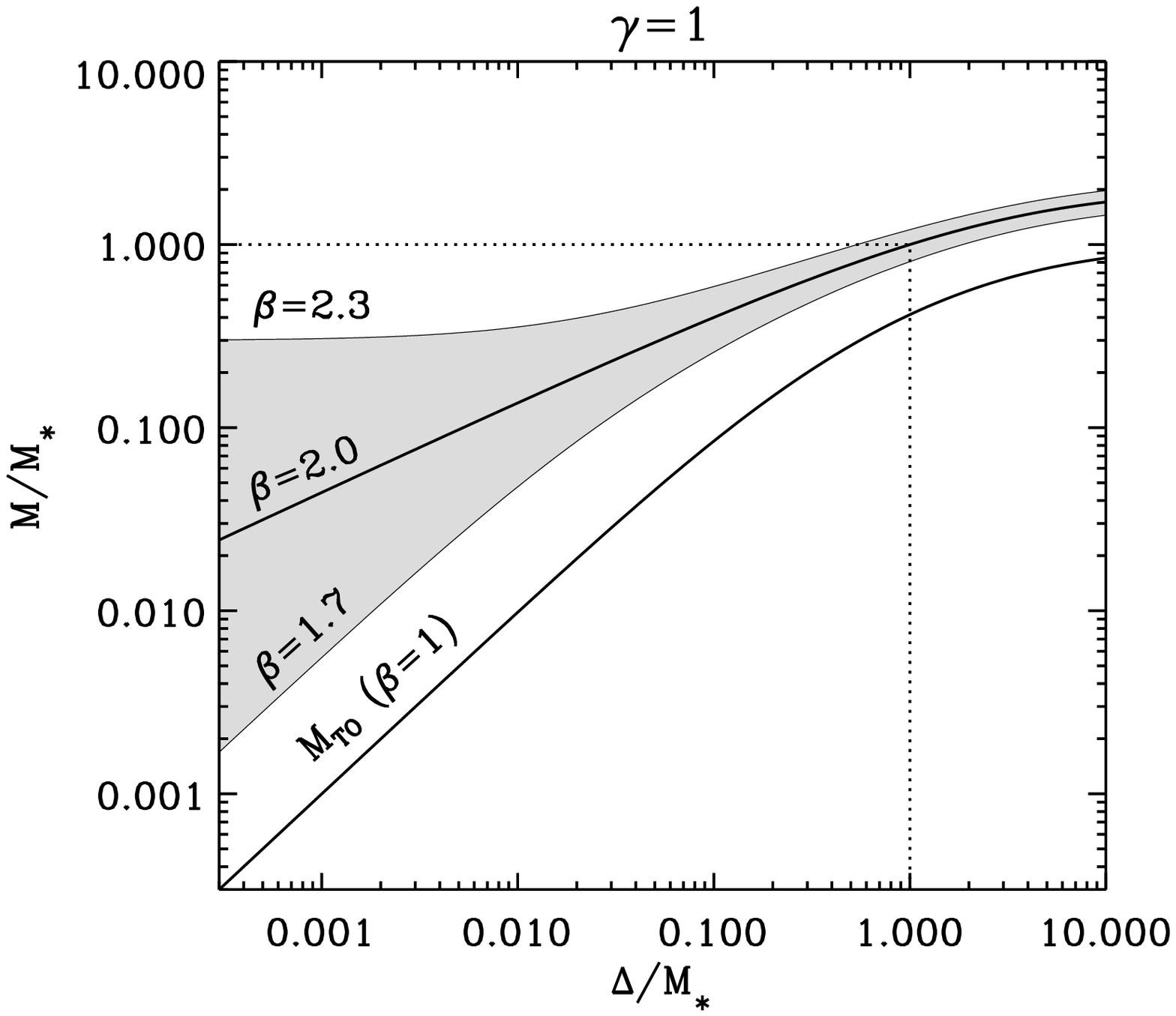} \includegraphics[width=8.cm]{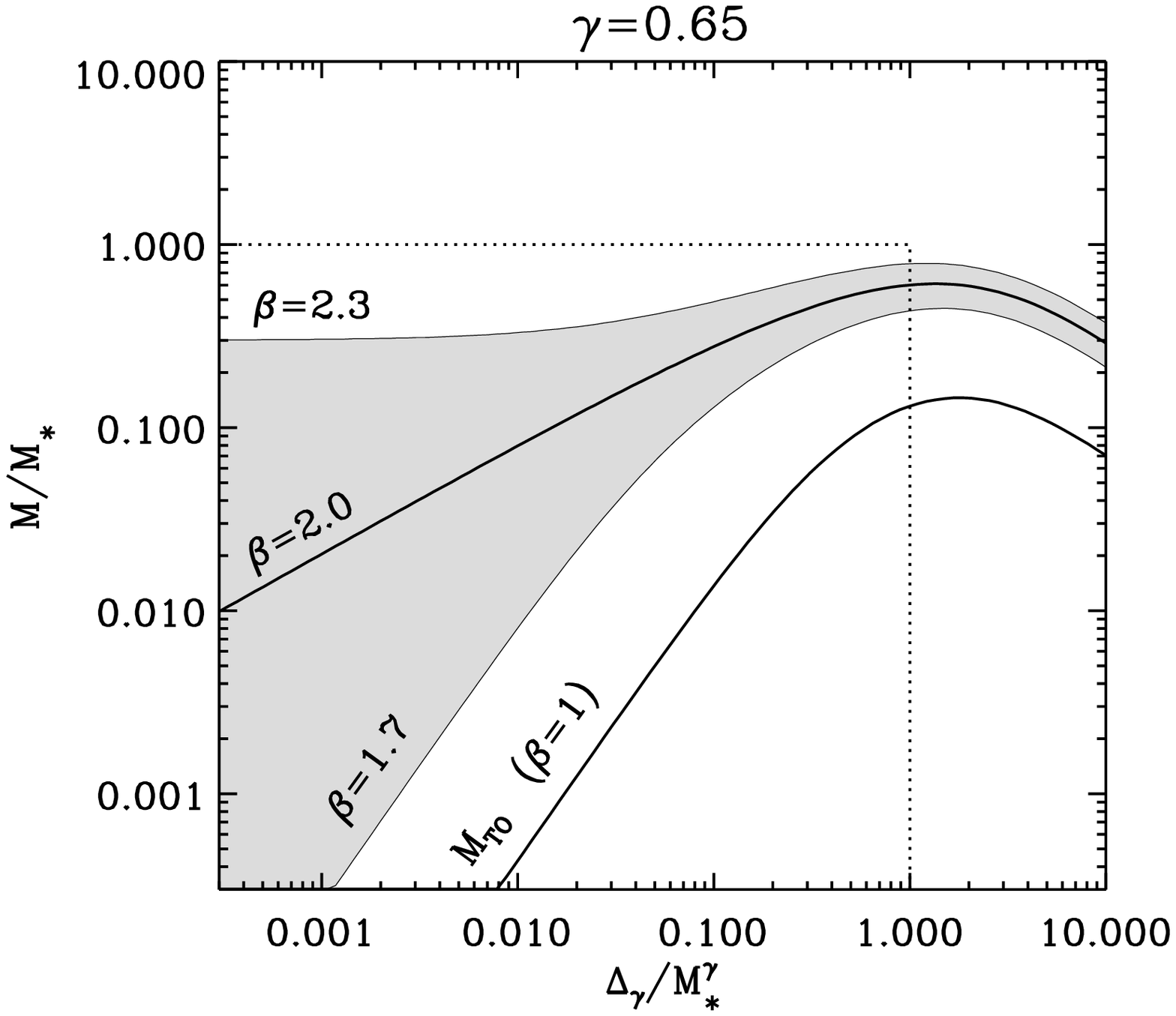}	
     \caption{The evolution with age of the mass where the evolved Schechter function has the logarithmic slope, $-\beta$, for different values of $\beta$. In the left panel a constant mass loss rate is considered (equation~\ref{eq:dndmjordan}, Section~\ref{subsec:lin}) and in the right panel the mass loss rate depends on mass  (equation~\ref{eq:dndmnonlin}, Section~\ref{subsec:nonlin}). 
     The $x$-axes represent  time or age since $\Delta\equiv t/t_0$ and $\Delta_\gamma\equiv\gamma t/t_0$.
     The mass for which $\beta=1$, i.e. the mass at which the mass function, presented as the number of clusters in logarithmic mass bins, turns over,
         is indicated as $\mto$. The grey shaded region indicates the part of the mass function where the logarithmic slope is $-2.0\pm0.3$.}
   \label{fig:slopes}
\end{figure*}

From equation~(\ref{eq:dndmjordan})  an expression for the  logarithmic slope (equation~\ref{eq:betadef}) can be derived
\begin{equation}
-\beta=-2+\frac{2\Delta}{M+\Delta}-\frac{M}{\mc}.
\label{eq:beta}
\end{equation}
The three terms on the right-hand side are, respectively, the initial index of the power law part of the Schechter function, a disruption term that makes the mass function shallower and a  truncation term that makes the mass function steeper. 
Solving for $M$ gives the mass at which the mass function has the logarithmic slope $-\beta$

\begin{equation}
M(\beta)\spaze=\spaze\frac{-\Delta\spaze+\spaze(\beta\spaze-\spaze2)\mc\spaze+\spaze\sqrt{(\Delta\spaze+\spaze(\beta-2)\mc)^2\spaze+\spaze8\Delta\mc}}{2}.
\label{eq:mbeta}
\end{equation}
From this it can be seen that $M(\beta)$ assumes its simplest form when $\beta=2$.  This is where the truncation term 
in equation~(\ref{eq:beta}) equals  the disruption term and this could, therefore, be interpreted as a point in the mass function where there is a balance between a steepening due to the truncation and a flattening by disruption.  Using equation~(\ref{eq:mbeta}) with $\beta=2$ an expression for $M(\beta=2)$ arises
\begin{equation}
M(\beta=2)=\frac{-\Delta+\sqrt{\Delta^2+8\Delta\mc}}{2}.
\label{eq:mba}
\end{equation}
From equation~(\ref{eq:mba}) it can be seen that in the limit of  $\Delta<<\mc$, i.e. at young ages where (massive) clusters are not yet affected by disruption, the scaling of $M(\beta=2)$ with $\Delta$ converges to  $M(\beta=2)=\sqrt{2\Delta\mc}$. Since $\Delta$ is a proxy for age, it means that the part of the mass function that can be approximated by a power law with index $-2$ is found at higher masses at older ages. This can also be seen  from the expression for $\beta$ as a function of $M$ and $\Delta$ in equation~(\ref{eq:beta}). For the mass function to have a logarithmic slope of $\beta=2$, the two right-hand terms (i.e. the disruption and truncation term) in equation~(\ref{eq:beta}) need to have a sum of 0, so $\Delta$ is higher at higher masses. 

From equation~(\ref{eq:mbeta}) it can be seen that in the same regime of $\Delta<<\mc$ the behaviour of $M(\beta\ne2)$  is different. For $\beta<2$ , $M(\beta)$ also increases with $\Delta$, but linearly, so faster  than $M(\beta=2)$. For $\beta>2$, $M(\beta)$ is constant. This is because at young ages the steep part of the mass function is not affected by disruption at all.

Equation~(\ref{eq:mbeta}) also allows us  to find the mass where the mass function, when presented as the number of clusters per logarithmic unit of mass, $\dndlogmtext$,  peaks or `turns over'. This is where $\beta=1$ and this $M$ is defined as the turnover mass, $\mto$. Its  dependence on aforementioned variables is (J07)
\begin{equation}
\mto=\frac{-(\mc+\Delta)+\sqrt{(\Delta+\mc)^2+4\Delta\mc}}{2}.
\label{eq:mto}
\end{equation}
For $\Delta<<\mc$ the turnover mass scales as $\mto\propto\Delta$, like $M(\beta<2)$, meaning that the turnover shifts to higher masses as time progresses. Notice that due to the different scaling of $M(\beta=2)$  and $\mto$ with $\Delta$,  they approach each other for $\Delta/\mc\gtrsim1$. In fact, for $\Delta>>\mc$ the shape of the mass function remains unchanged (J07) and approaches an equilibrium form  with $\mto=\mc$ and $M(\beta=2)=2\,\mc$. The number of clusters reduces as time progresses, but the shape of the evolved Schechter function remains the same. This means that clusters with initial masses much higher than $\mc$ are  replacing clusters with lower masses that have already been disrupted. Since the number of clusters above $\mc$ is very low, in practise it means that the total number of remaining clusters quickly drops to 0 after  $\Delta=\mc$.

The behaviour of the relations described above is illustrated in the left panel of Fig.~\ref{fig:slopes}. It shows $M$ vs. $\Delta$, with both quantities relative to $\mc$, such that the $x$ and $y$ scales are in dimensionless units of mass and time, respectively. The results for $\beta=1/1.7/2/2.3$ (equations~[\ref{eq:mbeta}], [\ref{eq:mba}]\,\&\,[\ref{eq:mto}]) are shown.

The grey shaded region in Fig.~\ref{fig:slopes} shows that there is a large region in the age vs. mass plane, where the logarithmic slope of the mass function equals $-2\pm0.3$.  This region  more or less coincides with  the strip that observed cluster populations occupy in the age-mass plane, implying that the observable part of the evolved Schechter function can be approximated by a power law with index $-2$. This will be verified in Section~\ref{sec:obs}  using  empirically derived ages and masses of star clusters in  M51.
 But first the case of a mass-dependent mass loss rate  is discussed in Section~\ref{subsec:nonlin}.

\subsection{Mass-dependent mass loss rate}
\label{subsec:nonlin}
In the previous section a (simplified) linear scaling between $\tdis$ and $\trh$ was adopted. However, there are theoretical arguments for a de-coupling  of $\tdis$ from $\trh$
for clusters dissolving in a tidal field \citep{2000MNRAS.318..753F}. \citet{2001MNRAS.325.1323B} predicted that  $\tdis\propto\trh^{3/4}$ for clusters that are initially  Roche-lobe filling. 
This leads to a simple scaling of $\tdis$ with  $M$ of the form  \citep{2005A&A...429..173L}
\begin{equation}
\tdis =t_0\,M^\gamma, 
\label{eq:tdis}
\end{equation}
with $t_0$ an environment dependent constant and $\gamma\simeq0.65$. 
This result for  Roche-lobe filling clusters was indeed found from $N$-body simulations of clusters dissolving in a tidal field \citep{1997MNRAS.289..898V, 2003MNRAS.340..227B} and it was also found for Roche-lobe underfilling clusters \citep{2008MNRAS.389L..28G}. The simple disruption law of equation~(\ref{eq:tdis}) was also used to describe age and mass distributions of clusters, from which a mean value of $\gamma=0.62\pm0.06$ was found (BL03).

The mass loss rate, $\mdot$, can be related to  $\tdis$ as
\begin{eqnarray}
\dmdt &= &\frac{M}{\tdis},\nonumber\\
           &=& \frac{M^{1-\gamma}}{t_0}.
\label{eq:mdot}
\end{eqnarray}
For $\gamma\ne1$ this results in a mass loss rate that is no longer constant, but instead becomes a function of $M$. Note that then $t_0$ is not simply $1/\mdot$ anymore and that $\tdis$ is no longer the time of total disruption as in Section~\ref{subsec:lin}. The total disruption time, i.e. the time it takes for all stars to become unbound, will be referred to  as $\tdistot$. The relation between $\tdistot$ and $\tdis$ will be derived a bit further down.

With $\tdis$ defined as in equation~(\ref{eq:tdis}) it is still possible to get an analytical expression for the mass evolution with time \citep{2005A&A...441..117L}
\begin{equation}
M=\left[M_i^\gamma- \Delta_\gamma\right]^{1/\gamma},
\label{eq:mtnonlin}
\end{equation}
where $\Delta_\gamma$ is defined as $\Delta_\gamma \equiv \gamma t/t_0$ to get a similar looking expression as in the constant mass loss case (equation~\ref{eq:mtlin}).
The effect of stellar evolution can be taken into account by replacing  $M_i^\gamma$ in equation~(\ref{eq:mtnonlin}) by $(\muev\,M_i)^\gamma$, where $\muev$ is a time-dependent  variable that represents the fraction of the initial stellar mass that has not been lost by stellar winds or super-nova explosions (typically between 0.7 and 1). 
For reasons of simplicity $\muev$ is not considered in the derivations that follow in this section. In  Section~\ref{subsec:full} the results including $\muev$ are given.
 The symbol $\Delta_\gamma$ is used when referring to $\gamma<1$ and $\Delta$ is used for the case of $\gamma=1$. The variable $\Delta_\gamma$ is a proxy of time, like $\Delta$, but note that $\Delta_\gamma$ does not represent the amount of mass that is lost from $M_i$, like $\Delta$, because of the powers of $\gamma$ and $1/\gamma$ in equation~(\ref{eq:mtnonlin}). The instantaneous  disruption time for a cluster with mass $M$ is  when $\Delta_\gamma/M^\gamma=1$ (equation~\ref{eq:mtnonlin}), such that $\tdistot(M)=(t_0/\gamma)\,M^\gamma$, or $\tdis=\gamma\tdistot$.  

The partial derivative that is needed (equation~\ref{eq:dndmdef})  to get an expression for the evolved Schechter function  can be found from equation~(\ref{eq:mtnonlin})
\begin{equation}
\dmidm = \left[1+ \frac{\Delta_\gamma}{M^\gamma}\right]^{1/\gamma-1},
\end{equation}
which combined with equation~(\ref{eq:dndmi})\,\&\,(\ref{eq:dndmdef}) gives an expression for the evolved Schechter function for a variable $\gamma$
\begin{equation}
\dndm = \frac{A\,M^{\gamma-1}}{\left[ M^{\gamma}\spaze+\spaze\Delta_\gamma\right]^{(\gamma+1)/\gamma}}\exp\left(\spaze-\frac{\left[M^\gamma\spaze+\Delta_\gamma\right]^{1/\gamma}}{\mc}\right).
\label{eq:dndmnonlin}
\end{equation}
For $M << \mc$ the exponential term in  equation~(\ref{eq:dndmnonlin}) is about 1 and the result for an nontruncated mass function of \citet{2005A&A...441..117L} is found,
which at low masses and old ages ($M^\gamma<<\Delta_\gamma$) goes to $\dndmtext\propto M^{\gamma-1}$. This is the consequence of the assumed relation between $\mdot$ and $\tdis$ (equation~\ref{eq:mdot}). If $\mdot$ was assumed to be constant in time, even when $\tdistot(M_i)\propto M_i^\gamma$, with $\gamma<1$, the low-mass end of the mass function would be flat (index of 0). The shape of the low-mass end of the evolved mass function is thus independent of the shape of the CIMF at  those masses and depends only on the way clusters lose mass \citep{2001ApJ...561..751F, 2005A&A...441..117L}.

Using the definition of the logarithmic slope, $-\beta$, from equation~(\ref{eq:betadef}) an expression for $\beta$ is found 
\begin{equation}
-\beta=-2+\frac{(\gamma+1)\Delta_\gamma}{M^\gamma+\Delta_\gamma} - \frac{M^\gamma\,\left(M^\gamma+\Delta_\gamma\right)^{1/\gamma-1}}{\mc}.
\label{eq:betanl}
\end{equation}

Unfortunately equation~(\ref{eq:betanl}) can only be solved for $M$ analytically for a few specific values of $\gamma$ and $\beta$ and the solutions are rather complicated. Therefore, $M(\beta)$ is solved numerically and  the right panel of Fig.~\ref{fig:slopes} shows the result for the same values of $\beta$ as in the constant mass loss case (left panel of Fig.~\ref{fig:slopes}) using $\gamma=0.65$.  For $M<<\mc$, $\mto$ scales as $\Delta_\gamma^{1/\gamma}$, which was also found under the assumption of a continuous power law in the instantaneous disruption  model  of  BL03. However, the increase of $\mto$ with age slows down at high ages, due to the truncation. In contrast to the constant mass loss case, the evolved Schechter function with $\gamma<1$ does not approach an equilibrium shape. For $\gamma=0.65$,  the value of $\mto$ reaches its highest  value ($\mto\simeq0.2\,\mc$) around $\Delta_\gamma\simeq2\mc^\gamma$ and then decreases again for older ages. This subtle difference might be important. If globular clusters formed from a Schechter type CIMF, with $\mc\simeq10^6\,\msun$, then together with the fact that $\mto\simeq2\times10^5\,\msun$, the constant mass loss model would say that the value of $\mto$ can still increase by roughly a factor of five (see left panel of Fig.~\ref{fig:slopes}). However, the $\gamma=0.65$ model shows that $\mto$ already has its highest possible value when $\mto/\mc\simeq0.2$ (right panel of Fig.~\ref{fig:slopes}). A possible explanation for the near universality of the value of $\mto$ could thus be that most globular cluster systems have reached already their maximum value of $\mto/\mc$ due to disruption and for a range of roughly an order of magnitude in $\Delta_\gamma$, the value of $\mto$ will be (within a factor of two) about $0.2\,\mc$. Since globular clusters are roughly coeval, a spread in $\Delta_\gamma$ can be interpreted as a spread in disruption time-scales, thereby allowing for a range of $\tdis$ values (i.e. different galactocentric distances) resulting in similar values of  $\mto$.  Perhaps this adds a piece to the puzzle of the problem of the universality of the globular cluster mass function (see also \citealt{1997ApJ...487..667O, 2000MNRAS.318..841V, 2002AJ....124..147W, 2008MNRAS.384.1231B}).

The mass for which the mass function has an index of $-2$ scales approximately as $\Delta_\gamma^{\eta}$, with $\eta\simeq0.6$, i.e. slightly different than the $\sqrt{\Delta}$ scaling found for the $\gamma=1$ case described in Section~\ref{subsec:lin}\footnote{In fact, by numerically solving equation~(\ref{eq:betanl}) for $\beta=2$ and  $0<\gamma\le1$ the relation between $\eta$ and $\gamma$ is $\eta\simeq1-0.85\,\gamma+0.35\,\gamma^2$.}. 
However, the $\Delta_\gamma^{0.6}$ scaling is a bit closer to the increase with age of the limiting mass due to the detection limit, $\mlim(\tau)$, of empirically derived cluster masses. For a $V$-band detection limit, $\mlim(\tau)$ scales as $\tau^{0.7}$.  {\it This implies that the observable part of the evolved Schechter function, i.e. above the detection limit, has  a logarithmic slope of approximately $-2$ at all ages.} 
This is an important result, because it means that there is no significant difference to be expected between the $\cimfe$ and the shape of the cluster mass function at old ages. So the argument that disruption needs to be mass-independent because old clusters have a similar mass function as young clusters does not hold when a Schechter function is assumed for the CIMF.

All results presented for the constant mass loss case in Section~\ref{subsec:lin} are simply a subset of the more general solutions presented in this section. That is, when using $\gamma=1$, $\Delta_\gamma=\Delta$ and equations~(\ref{eq:tdis}), (\ref{eq:mtnonlin}), (\ref{eq:dndmnonlin})\,\&\,(\ref{eq:betanl})  from this section are the same as equations~(\ref{eq:tdislin}), (\ref{eq:mtlin}), (\ref{eq:dndmjordan})\,\&\,(\ref{eq:beta}), respectively, from Section~\ref{subsec:lin}.  For completeness a general expression for  the evolved Schechter function, including mass loss by stellar evolution and a variable initial power law index, is presented in Section~\ref{subsec:full}. 

\subsection{Including stellar evolution and a variable power law index}
\label{subsec:full}
As mentioned in the beginning of this section all derivations are done for a fixed power law index of $-2$ (equation~\ref{eq:dndmi}) and without the inclusion of mass loss due to stellar evolution. The expressions for the evolved Schechter function and the logarithmic slope $-\beta$ are here given for a variable initial power law index,  $-\alpha$, and an additional term $\muev$, which is the fraction of the original mass that is not lost  due to stellar evolution. Therefore, $\muev$ is a function of time and can be taken from an SSP model. See also \citet{2005A&A...441..117L} for analytical approximations of $\muev(t)$.
Following the same steps as in Section~\ref{subsec:nonlin} the expressions for $\dndmtext$ and $\beta$ are

\begin{equation}
\dndm = \frac{A\,\muev^{\alpha-1} M^{\gamma-1}}{\left[ M^{\gamma}\spaze+\spaze\Delta_\gamma\right]^{(\gamma+\alpha-1)/\gamma}}\exp\left(\spaze-\frac{\left[M^\gamma\spaze+\Delta_\gamma\right]^{1/\gamma}}{\muev\mc}\right),
\label{eq:dndmfull}
\end{equation}
and 

\begin{equation}
-\beta=-\alpha+\frac{(\gamma+\alpha-1)\Delta_\gamma}{M^\gamma+\Delta_\gamma} - \frac{M^\gamma\,\left(M^\gamma+\Delta_\gamma\right)^{1/\gamma-1}}{\muev\mc}.
\label{eq:betafull}
\end{equation}

From equations~(\ref{eq:dndmfull})\,\&\,(\ref{eq:betafull}) it can be seen that $\muev$ does not affect the shape of the mass function at low masses, it only affects that vertical offset by a bit. The exponential truncation occurs at slightly lower masses, at $\muev\mc$ instead of $\mc$.

\section{Application to M51}
\label{sec:obs}

To illustrate the cluster population model of the previous section,  some of its elements are compared  to the cluster population of  the interacting galaxy M51 (NGC~5194). The cluster ages and masses were determined from HST/WFPC2 multi-band photometry by Bastian~et~Êal.~(\citeyear{2005A&A...431..905B}, hereafter B05). The masses are corrected for mass loss by stellar evolution, i.e. they represent the sum of the initial masses of the stars that are still bound in the clusters. The present day masses of the oldest clusters are, therefore,  $\sim$25\% lower. The M51 cluster population is a good benchmark to test the evolved Schechter function for several reasons: firstly, it is a rich cluster population hosting several massive star clusters   (\citealt{2000MNRAS.319..893L}; B05; \citealt{2005AJ....130.2128L, 2008AJ....135.1567H}).   M51 hosts   sufficient clusters above $\sim$10$^4\,\msun$, which is the approximate minimum mass for the optical fluxes not to be affected  by stochastic effects due to sampling of the stellar IMF, allowing reliable age dating through multi-band photometry (e.g. \citealt{2004A&A...413..145C}).
Secondly, for this cluster population it has been suggested that the mass function is truncated \citep{2006A&A...446L...9G, 2008A&A...487..937H}. Lastly, the disruption time of clusters in M51 is thought to be very short (BL03, \citealt{2005A&A...441..949G}), due to the strong tidal field  and the high density of GMCs in the inner region of the galaxy, where most of the clusters of the B05 sample reside. These environmental properties suggest a noticeable  evolution of the mass function. Before a comparison between the model and the data is given, the parameters for the CIMF ($\mc$) and the disruption law (equation~\ref{eq:tdis}) are determined in Section~\ref{subsec:maxlike}.

\subsection{A fit of the evolved Schechter function to the M51 cluster population}
\label{subsec:maxlike}
If a constant cluster formation history is assumed the evolved Schechter function (equation~\ref{eq:dndmnonlin})  essentially  represents  the probability of finding a cluster with mass between $M$ and $M+\dr M$ and age between $\tau$ and $\tau+\dr \tau$. It can thus be used as a two-dimensional distribution function to determine all the parameters of the evolved Schechter function from empirically derived ages and masses using a maximum likelihood estimate. The only thing which needs to be added is the detection limit. For this the result of the incompleteness analysis of B05 is used, who show that their cluster sample is limited by a detection in the F435W (roughly $B$) band, for which the 90\% completeness fraction is a 22.6 mag.  The same simple stellar population (SSP) model  was used as in B05 to derive the ages and mass, namely the GALEV models for Salpeter stellar IMF between $0.15\,\msun$ and $50\,\msun$ \citep{2002A&A...392....1S, 2003A&A...401.1063A}. The photometric evolution of a single mass cluster in the F435W band from the SSP model, combined with the 90\% completeness limit and the distance to M51 (8.4\,Mpc, \citealt{1997ApJ...479..231F}) then gives the limiting cluster mass as a function of age, $\mlim(\tau)$ (BL03; \citealt{2007ApJ...668..268G}). This limiting mass, together with the cluster ages and masses, is shown in Fig.~\ref{fig:slopem51} and in panel~(a) of Fig.~\ref{fig:obs}.

Using $\mlim(\tau)$ and equation~(\ref{eq:dndmnonlin}) artificial models for varying $\gamma$, $t_0$ and $\mc$ are built and a (simultaneous) maximisation of the likelihood of these parameters gives

\begin{itemize}
\item $\mc=(1.86\pm0.52)\times10^5\,\msun$;
\item $t_0=0.19\pm0.10\,$Myr;
\item $\gamma=0.67\pm0.06$.
\end{itemize}
The (statistical) uncertainties on each variable are determined using a bootstrap method.   For this the original data-set is randomly re-sampled 1000 times, allowing for multiple entries of the same age-mass pair and omission of values, such that the total number of age-mass pairs is the same. The likelihood of the three parameters is determined for each of these 1000 samples and the standard deviation of the results for each parameter is used as the uncertainty.   

The disruption parameters ($ t_0$ and $\gamma$) imply a total dissolution time for a cluster with an initial mass $\mc$ of $\tdismc=(t_0/\gamma)\,\mc^\gamma\simeq950\,$Myr, implying that there should not be many clusters that survive longer than a Gyr. Indeed B05 found only a handful of clusters older than 1 Gyr, although this is probably not only caused by disruption, but also by detection incompleteness.
A cluster with an initial mass of $10^4\,\msun$ gets completely destroyed in $\tdisfour\simeq130\,$Myr.  This latter value is in perfect agreement with the results of  \citet{2005A&A...441..949G}, who found $100\lesssim\tdisfour\lesssim200$, depending on what is assumed for the cluster formation history. This very short lifetime of clusters in M51, as compared to the lifetime of clusters in the solar neighbourhood or the Magellanic Clouds, was attributed to the high molecular cloud density in this galaxy \citep{2006MNRAS.371..793G}.

The value of $\gamma=0.67\pm0.06$ agrees perfectly with the mean result of BL03 who found $\gamma=0.62\pm0.06$ from fits to the age and mass distributions of clusters in several galaxies. This value also follows from theory and $N$-body simulations of clusters dissolving in a tidal field, since for these clusters the disruption time-scale can be expressed as $\tdis\propto(N/\log\Lambda)^{0.75}$ \citep{2001MNRAS.325.1323B, 2003MNRAS.340..227B, 2008MNRAS.389L..28G}. In this equation, $\log\Lambda\simeq\log(0.1N)$ is the Coulomb logarithm and it follows that this expression for $\tdis$ can be well approximated by $\tdis\propto N^{0.62}$ when a relevant range of $N$ is considered  \citep{2005A&A...429..173L}. The disruption time due to external perturbations scales (on average) in a similar way with mass.  It actually scales linearly with the density of clusters, but since the radius of clusters depends only weakly on their mass, $\tdis$ due to external perturbations (GMCs, spiral arms, etc.) also scales as $M^\gamma$ with $\gamma$ slightly smaller than 1 \citep{2006MNRAS.371..793G, 2006A&A...455L..17L}. \citet{2005A&A...441..949G} also looked at the mass dependence of cluster disruption in M51 and found that the value of $\gamma$ depends on the mass range considered. When excluding the most massive clusters ($M\gtrsim 10^5\,\msun$) a value of $\gamma=0.6$ was found. Here it is shown that, when considering a Schechter function for the CIMF, the full cluster population can be fit with a disruption law in which $\tdis$ depends on mass. 

The value of $\mc$ is in excellent agreement with the result of \citet{larsen08} who found $\mc=(2.0\pm0.5)\times10^5\,\msun$ for a sample of spiral galaxies. 
\citet{2006A&A...446L...9G} found a slightly lower value of $\mc=10^5\,\msun$ through a model comparison to the LF of M51 clusters. Since the LF is usually constructed for apparent luminosities, i.e. not corrected for local extinction, the derived value of  $\mc$ from the LF can be underestimating the true value of $\mc$. \citet{2006A&A...446L...9G} applied a correction of $A_V=0.25\,$mag to all clusters to roughly account for extinction, but B05 showed that the youngest clusters, which are the brightest ones,  are slightly more extincted than this. Extinction of the brightest clusters  has a  similar effect on the LF  as lowering $\mc$, so this could be why the value for $\mc$ found by \citet{2006A&A...446L...9G} is slightly lower than that found here.

\subsection{Logarithmic slopes at different ages}
\begin{figure}
 \includegraphics[width=8.cm]{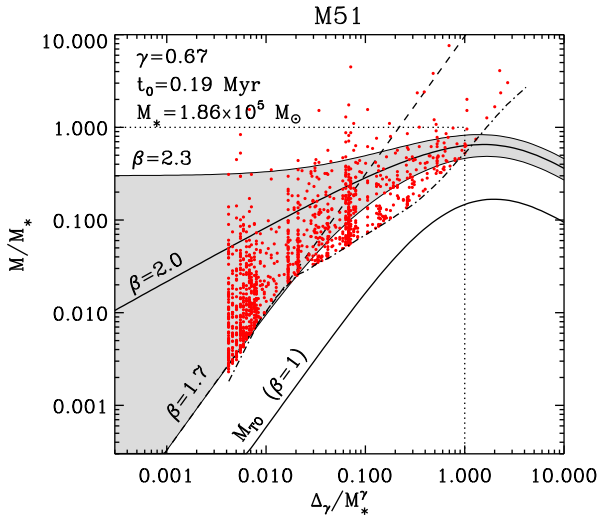} 
     \caption{The time evolution of the mass where the evolved Schechter function has the logarithmic slope, $-\beta$, for $\beta=1, 1.7, 2$ and 2.3 (same as in the right panel of Fig.~\ref{fig:slopes}).   The dots are ages and masses of M51 clusters from B05, with the dot-dashed line defining  the limiting mass due to a detection limit of 22.6 mag in the F435W filter. The dashed line shows $M(\beta=1.7)$ when a continuous power law CIMF is assumed. The values of $\gamma, t_0$ and $\mc$, used to normalise the ages and masses to dimensionless units, follow from a maximum likelihood estimation (Section~\ref{subsec:maxlike}).}
   \label{fig:slopem51}
\end{figure}

Fig.~\ref{fig:slopem51}  shows the M51 ages and masses in dimensionless units on top of the predictions for the evolution of several values of $\beta$. The results of the maximum likelihood estimation from Section~\ref{subsec:maxlike} are used for the  calculation of $M(\beta)$ and the normalisation of age and mass.

 The increase of the minimum observable cluster mass, $\mlim(\tau)$, is shown as a dot-dashed line marking the lower envelope of the data points. Most of the clusters fall in the grey region, for which the predicted logarithmic slope of the mass function is $-2\pm0.3$, with the $M(\beta=2)$ line  showing a very similar increase with age as the data points.
  At old ages $\mlim(\tau)$ approaches the $M(\beta=2)$ line, implying that the mass function at these ages is somewhat steeper than a $-2$ power law. Although the disruption time of clusters in M51 is very short, $\mto$ does not get closer than roughly one mass dex below $\mlim(\tau)$ and is, therefore, not observable at any age. The dashed line in Fig.~\ref{fig:slopem51} shows $M(\beta=1.7)$ for  the case of a continuous power law CIMF (BL03). This line shows that disruption would clearly make the mass function of the oldest half of the data flatter than a $-2$ power law if the CIMF was a continuous power law with that index,    whereas the  $M(\beta=1.7)$ line for the evolved Schechter function bends down at old ages and stays (almost) below the detection limit.
          
A direct comparison between the evolved Schechter function (Section~\ref{subsec:nonlin}) and the M51  cluster mass functions at different ages  is given in Section~\ref{subsec:dndtdndm}.

\subsection{Mass functions at different ages}
\label{subsec:dndtdndm}
\begin{figure*}
\center\includegraphics[width=17.cm]{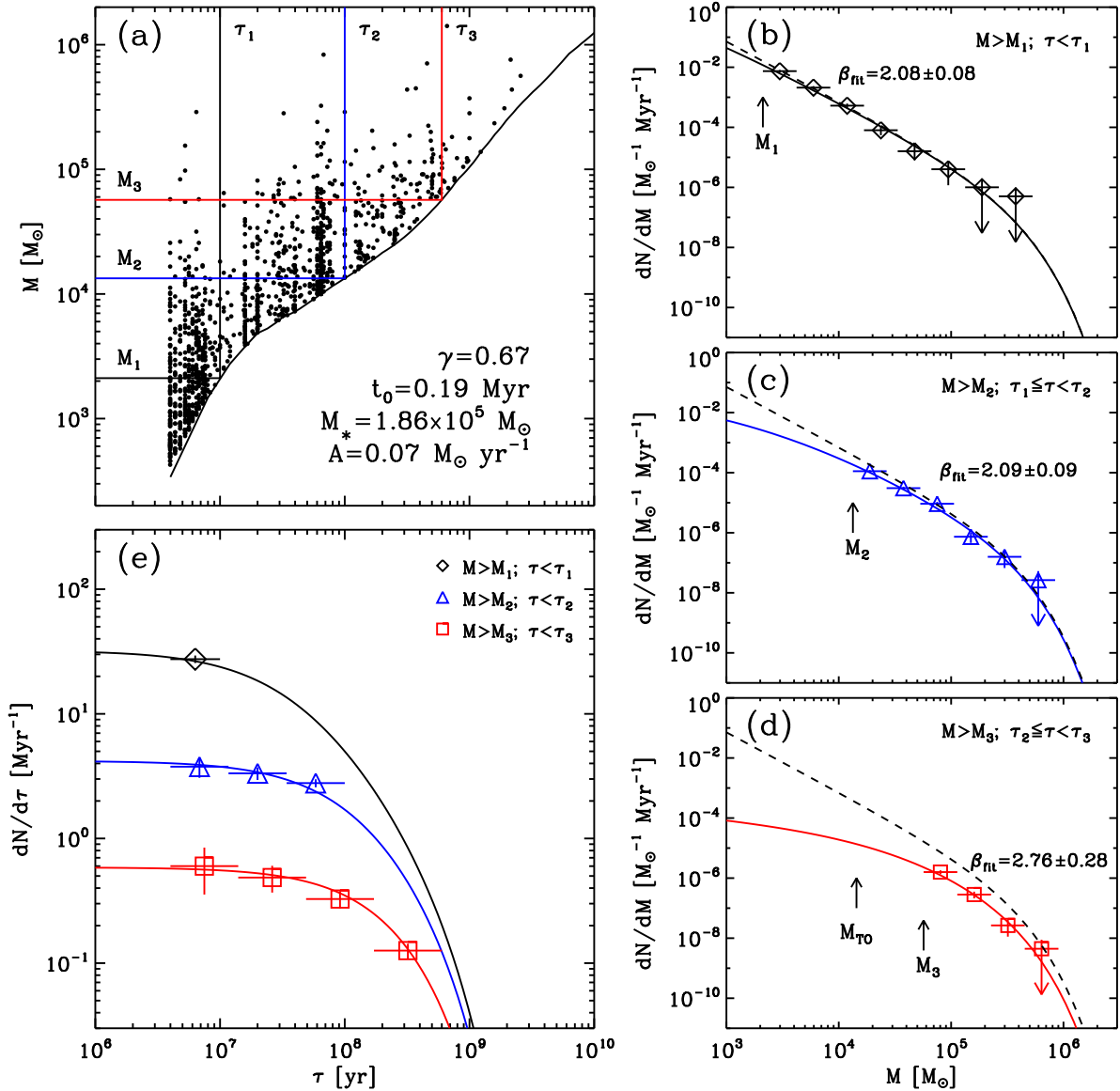} 	
     \caption{Ages and masses of M51 clusters (panel [a]) and the resulting mass functions in different age bins  (panels~[b]-[d]) with the evolved Schechter  functions (equation~\ref{eq:dndmnonlin})  over-plotted. The parameters were determined by a maximum likelihood fit of the model to the data (Section~\ref{subsec:maxlike}). The empirical mass functions are approximated by power laws and the resulting index, $-\betafit$, is indicated in each panel.
The  mass limited   age distributions for different mass cuts are shown in panel (e).  The vertical positioning of the age distributions with respect to one another is due the fact that
       each sample has a different lower mass limit ($M_j$). All mass functions and age distributions are described by a single value of $A=0.07\,\msunyr$, corresponding to $\Gamma\simeq0.1$, i.e. 10\% of the star formation in M51 occurs in star clusters.
       }
   \label{fig:obs}
\end{figure*}

The presentation of cluster ages and masses in dimensionless units in the previous section is not very common. Therefore,  the age-mass diagram is presented again, but now in physical units (panel~[a] of Fig.~\ref{fig:obs}). Three age bins are created, with boundaries $4, 10, 100$\,Myr and 600\,Myr. The upper boundaries are indicated by $\tau_1, \tau_2$ and $\tau_3$, respectively. Due to the increasing $\mlim(\tau)$ with $\tau$, three corresponding mass limits are defined as $M_j=\mlim(\tau_j)$, with $1<j<3$. The values of $M_j$ are roughly $2\times10^3\,\msun, 10^4\,\msun$ and $6\times10^4\,\msun$.

 The empirical mass functions in different age bins are shown in panels~(b)-(d) of Fig.~\ref{fig:obs}.  In each panel an arrow denotes the lower mass limit of each sample, $M_j$. The first  mass bin starts at this lower limit  and the number of clusters in each mass bin is counted and then divided by the width of the bin, such that $\dndmtext$ is obtained. The $\dndmtext$ is also divided by the age range of the sample. In this way the histogram points represent the number of clusters per unit of mass and time/age, or $\dr N/(\dr M\dr \tau)$, which can be compared to the evolved Schechter function. The variables for $\gamma$, $t_0$ and $\mc$ as determined in Section~\ref{subsec:maxlike} are used for the model predictions.
 
The dashed lines show the CIMFs (equation~\ref{eq:dndmi}) and the evolved Schechter functions (full lines, equation~\ref{eq:dndmnonlin}) are based on the mean age of the M51 clusters in each age bin, being approximately $5\,$Myr, 50\,Myr and $250$\,Myr for the panels~(b), (c) and (d), respectively. The vertical offset is determined by $A$, and the best agreement is found for $A=0.07\,\msunyr$. The coincidence of the empirical mass functions with the CIMF (dashed line) in the first two age bins,  shows that the number of clusters per linear unit of time is approximately constant in the first $100\,$Myr. The implications of this will be discussed in more detail in Section~\ref{subsec:dndtdndt}.  Only the $\dndmtext$ in the oldest age bin falls visibly below the CIMF. If a continuous power law distribution function would have been assumed for the CIMF, many more high-mass clusters would have to have been destroyed to make the model prediction agree with the data (already noted by Gieles~et~al.~[\citeyear{2005A&A...441..949G}] who do not include the most massive clusters when determining the disruption time of M51 clusters). The evolved Schechter functions describe the empirically derived mass functions very well and as could be seen already in Fig.~\ref{fig:slopem51}, the turnover of the evolved Schechter functions remains well below the detection limit.
 
 The empirical mass functions are also approximated by power laws and their indices, $-\betafit$, are indicated in each panel. The empirical mass functions  in the first two age bins can be approximated by a power law with index of roughly $-2$, but the oldest mass function is steeper. From a comparison to the (evolved) Schechter function, we see this is because  the limiting mass approaches $\mc$ at this age, where $\beta=3$. On the low-mass end of the evolved Schechter function the effect of mass-dependent disruption is seen at old ages (panel~[d]), where the mass function flattens to a logarithmic slope of $\gamma-1=-0.33$ and the turnover is at $\mto\simeq10^4\,\msun$. 
 
 Though this is only an application to one data-set, it shows that a Schechter function evolved with mass-dependent disruption nicely describes  empirically derived age and mass distributions. {\it The lack of a flattening of the  empirical mass function at old ages, as compared to the CIMF$_{\textup{\scriptsize{emp}}}$, is due the exponential truncation at high masses and the detection limit that shifts the mean observed masses closer to $\mc$, where the CIMF is steeper.}

\subsection{Age distributions for different mass cuts}
\label{subsec:dndtdndt}
Assuming  a constant formation history of clusters, the cluster age distribution, $\dndttext$, can be found  from a numerical integration of the evolved Schechter function from $M_j$ to $\infty$ at each $\tau$.

In panel~(e) three (mass limited) age distributions are shown, for the three maximum ages, $\tau_j$, and corresponding lower mass limit, $M_j$. The clusters are binned, such that the first bin starts at 4\,Myr, the age of the youngest cluster, and the last bin ends at $\tau_j$. The number of bins is chosen such that the bin widths are approximately constant for the three mass cuts. The number of clusters in each bin is counted and divided by the bin width, such that  $\dndttext$ follows, i.e. the number of clusters per Myr. 

Over-plotted as full lines are the  age distribution resulting from the numerical integrations of the evolved Schechter functions again using the best-fitting parameters from the maximum likelihood estimate of Section~\ref{subsec:maxlike}.
A single value of $A=0.07\,\msunyr$, i.e. the same value as in Section~\ref{subsec:dndtdndm}, was used to describe all three age distributions. The different vertical offsets are caused by the different values of $M_j$. The good agreement between the model and the three empirical age distribution shows that both the disruption parameters and the CIMF shape, with the parameters
 from the model fit of Section~\ref{subsec:maxlike}, provide a good description of the cluster population of M51.
With the value of $A$ and the global $\sfr$ of M51 it is possible to derive $\Gamma$, i.e. the fraction of star formation occurring in star clusters that survive the embedded phase ($\cfr=\Gamma\cdot\sfr$, \citealt{bastian08}). The global  $\sfr$ of M51 is around $5\,\msunyr$ (e.g. \citealt{2001AJ....122.3017S}). 
Since $\Gamma\cdot\sfr=A\,{\textup E}_1(\mmin/\mc)$ (equation~\ref{eq:mtot00}), with $\textup{E}_1(\mmin/\mc)$ roughly between  6 and 10, depending on $\mmin$, the value of $\Gamma$ resulting from this comparison is then $\Gamma\simeq0.08-0.14$.
 So roughly 10\% of the star formation in M51 occurs in star clusters that are detectable in the optical. This is in good agreement with the general finding of \citet{bastian08} that $\Gamma=0.08\pm0.03$ in different  galactic environments. 

The theoretical age distributions  all show an initially flat part and then a rapid decline. For the lowest mass cut ($M_1$) the data do not allow us to verify this, due to the young age at which the $\mlim(\tau)$ line cuts the sample ($\tau_1=10\,$Myr). The $\dndttext$ of the sample with the highest mass cut nicely shows this flat part and the decline.
The solutions for the mass limited age distributions can be approximated within $\sim$10\% accuracy by 
\begin{equation}
\dndt(M\spaze>\spaze M_j)\simeq\spaze\frac{A}{\mc}\spaze\left(\spaze\frac{\tdismc}{\tau\spaze+\spaze\tdismi}\spaze\right)^{\spaze\spaze1/\gamma}\hspace{-0.2cm}{\textup E}_2\spaze\spaze\left(\frac{M_j}{\mc}\right)\exp\left(-2\tau/\tdismc\right)
\label{eq:dndtap}
\end{equation}
where $\tdismc$ is the total disruption time of a cluster with an initial mass $\mc$ and $\tdismi$ is the same for a cluster with an initial mass $M_j$. This approximation only holds  for an initial power law index of $-2$ at low masses. The approximation  in equation~(\ref{eq:dndtap}) has the $\tau^{-1/\gamma}$ scaling predicted by BL03 for a continuous power law mass function, but it falls off faster (exponentially) at old ages, due to the truncation in the CIMF. 
 
 A flat part in $\dndttext$ at young ages is a typical result of the MDD model (BL03), although BL03 present the age distributions for luminosity limited cluster samples, which decline at young ages as $\tau^{-\zeta}$ due to evolutionary fading of clusters. 
 In the prediction for $\dndttext$ in this study the evolutionary fading does not enter because for each age distribution  a lower mass limit  is adopted that is above the fading line (Fig.~\ref{fig:obs}).  A  flat $\dndttext$ was also found for a mass limited sub-sample of clusters in the SMC up to  almost a Gyr \citep{2007ApJ...668..268G}. Indirect evidence for a flat $\dndttext$ for clusters in different galaxies follows from the linear scaling of $\mmaxlt$ with $\tau$  in the first 100\,Myr (Section~\ref{subsubsec:ev} and \citealt{2008A&A...482..165G}).
   
Note that there is no evidence for `infant mortality'  from the $\dndttext$ in the first $\sim$10\,Myr, as was reported by B05 from the same data set. B05 used slightly smaller age bins and found evidence for a drop of a factor of four  between the first age bin ($\lesssim10\,$Myr) and the second age bin. They also report an enhancement of clusters around 60\,Myr, which coincides with the moment of the last encounter between M51 and the companion galaxy (NGC~5195). However, there is  an artificial age gap between 10\,Myr and 20\,Myr  in the $\dndttext$, which is a common feature  when broad-band photometry is used to derive ages \citep{2005A&A...441..949G, 2005AJ....130.2128L, 2007AJ....133.1067W}. These three effects average out when using slightly larger age bins and the details in the $\dndttext$ reported by B05 are not separable anymore. The difference between the first and the second bin for the sample with the highest mass cut is within $1\,\sigma$ still consistent with a decrease of a  factor of two or three. However, panel~(e) of Fig.~\ref{fig:obs}  shows that a flat $\dndttext$ up to $\sim$$100$\,Myr describes
the data very well. 

\begin{figure}
\center\includegraphics[width=8.cm]{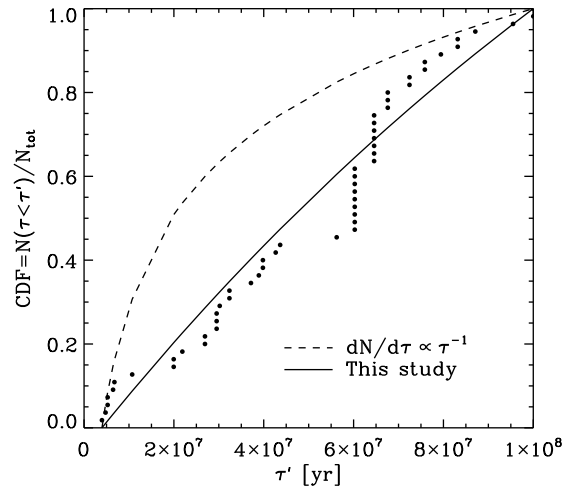} 	
     \caption{Cumulative distribution function (CDF) of  ages ($\le100\,$Myr) for the most massive clusters ($M\ge M_3\simeq6\times10^4\,\msun$). The dashed and full lines show the CDF for the 90\% MID model of \citet{2007AJ....133.1067W} and the one presented in this study, respectively. 
     }
   \label{fig:ks}
\end{figure}

Since this is  a quite different interpretation of the data than what was concluded by B05,  caused by a difference in how the data are binned, it is interesting to have a closer look at  the $\dndttext$ of massive clusters with a method that does not rely on binning.  One way of doing that is by creating a cumulative distribution function (CDF), which can then be compared to different models using a Kolmogorov-Smirnoff (K-S) test. 
 In Fig.~\ref{fig:ks}  the CDF of the first 100\,Myr of the $\dndttext$ of clusters more massive than $M_3\simeq6\times10^4\,\msun$ (Fig.\ref{fig:obs})  is shown (dots). The best-fitting model from Section~\ref{subsec:maxlike} is shown as a full line. The null hypothesis that the  data have been drawn from this model cannot be rejected. The K-S probability is 10\%, i.e. the model is within 2$\sigma$ consistent with these data.

\citet{2007AJ....133.1067W} claim that cluster evolution is  `universal'  in the first 100\,Myr and that the fraction of disrupted clusters in this period is independent of environment and mass.  In their model,  90\% of the clusters gets destroyed each age dex,  resulting in a $\tau^{-1}$ scaling of the $\dndttext$ for mass limited samples. The CDF of this model is shown as a dashed line in Fig.~\ref{fig:ks}.  The K-S probability for the Whitmore et al. model is $2\times10^{-8}$ and the hypothesis that these data are drawn from a $\tau^{-1}$ age distribution can, therefore, 
be safely reject . Even if the  disruption fraction is lowered to 80\%(70\%), resulting in $\dndttext\propto\tau^{-0.7}(\tau^{-0.5})$ \citep{2007AJ....133.1067W}, the K-S probability is $6\times10^{-5}(10^{-3})$. 

The fact that the full optically selected cluster population of M51 can be described by a CFR that is roughly 10\% of the SFR indicates that {\it if} all stars form in embedded clusters, 90\% of them are destroyed when the residual gas of the star formation process is removed and that this process lasts only a few Myrs. 
And it seems that `infant mortality'  does not affect the optically detected clusters and that information on the number of embedded clusters is needed to estimate the `infant mortality rate',  as was also done for the clusters in the solar neighbourhood \citep{2003ARA&A..41...57L}.

\section{Conclusions and discussion}
\label{sec:conclusions}
This study considers the early evolution (first $\sim$Gyr) of the star cluster mass function, with particular focus on the mass range that is available through observations.  A Schechter type function, i.e. a power law with an exponential truncation (equation~\ref{eq:dndmi}), is used for the cluster initial mass function (CIMF). The use of this function is motivated by observational indications that the high-mass end of the CIMF of young extra-galactic clusters is steeper than the `canonical'  $-2$ power law \citep{2006A&A...450..129G, larsen08}. The exponential truncation provides a good description of the high-mass tail of the globular cluster mass function (e.g. \citealt{1996ApJ...457..578M}).

Empirical cluster masses are generally found to be higher at older (logarithmic) ages. This is an observational bias because of two effects: the rapid fading of star clusters with age, making it increasingly more difficult to see low-mass clusters  with increasing age; also, masses increase due to a size-of-sample effect, since longer time intervals are considered at higher logarithmic ages. If the CIMF has indeed an exponential truncation, this means that at older (logarithmic) ages the steepening is more noticeable than at young ages.

The shape of the cluster mass function at different ages depends on the functional form of the CIMF and the way clusters lose mass due to disruptive effects.
Two competing models exist for the evolution of clusters, a mass-dependent disruption model (MDD, e.g. \citealt{2003MNRAS.338..717B}) and a mass-independent disruption model  (MID, e.g. \citealt{2007AJ....133.1067W}) and both assume that star cluster masses are drawn from a continuous power law distribution with index $-2$.

In this contribution a Schechter function is used for the CIMF, with a power law index of $-2$ at low masses and an exponential truncation at $\mc$ (equation~\ref{eq:dndmi}), which is evolved with mass-dependent cluster disruption. 
In summary, it is found that

\begin{itemize}
\item the exponential truncation of the Schechter function is not necessarily detectable from a small cluster sample,  the kind that is used to create the mass function of clusters younger than $\sim$10\,Myr (here referred to as $\cimfe$), because in this short time interval there are generally not enough clusters sampled above $\mc$;
\item the age distribution of mass limited sub-samples is flat during the first $\sim$100\,Myr (depending on the mass cut and the disruption time-scale) and drops exponentially at older ages. Through a comparison to ages and masses of  clusters in M51, it is shown that this holds for a sub-sample of clusters more massive than $6\times10^4\,\msun$. The $\tau^{-1}$ age distribution, proposed by \citet{2007AJ....133.1067W} as the result of a `universal' cluster disruption model, is ruled out at high significance for these data;
\item  the mass for which the logarithmic slope of the evolved Schechter function is  $-2$ increases with age as $\tau^{0.6}$. This scaling is similar to the increase of the limiting cluster mass, $\mlim(\tau)$, due to the evolutionary fading of clusters. This means that the mass function of clusters above the detection limit is  approximately a power law with index $-2$ at all ages.
\end{itemize}

The MDD model, based on a continuous (nontruncated) power law CIMF, predicts that the mass function at old ages should have a logarithmic slope of $\gamma-1\simeq-0.35$ \citep{2005A&A...441..117L}, while empirically derived  mass functions at old ages are power laws with indices of approximately $-2$, or even steeper (e.g.~\citealt{2006MNRAS.366..295D}). This weakness of the MDD model was at the same time a supporting argument for the MID model. However, in this study it shown  that a power law mass function with index $-2$ at old ages results naturally when a Schechter type CIMF is assumed. This is illustrated in  Section~\ref{sec:obs}, where it is shown that a Schechter function evolved with mass-dependent disruption provides an excellent description of the (mass limited) age distribution and the mass function at different ages for star clusters in M51. The power law mass function of M51 clusters gets steeper with age, from an index of $-2$ at the youngest ages to an index of $-2.8$ at $\sim$250\,Myr. However, simultaneous fitting of  models with different CIMF and disruption parameters to the ages and masses shows that the disruption time depends on mass as $\tdis\propto M^{0.67\pm0.06}$. This means that the logarithmic slope of the CIMF that is observable at $\sim$250\,Myr is smaller than $-2.8$ (i.e. steeper), since it has already been affected by disruption.

It would be interesting to apply the evolved Schechter function to more cluster populations for which age and mass information is available. As demonstrated in Section~\ref{subsec:maxlike}, the fundamental parameters defining a cluster population, namely $\mc$ 
 of the CIMF  and $t_0$ and $\gamma$ of the disruption law (equation~\ref{eq:tdis}) can easily be determined from luminosity limited cluster samples using a maximum likelihood estimate.
 \citet{2008MNRAS.383.1103P} apply the MDD model to the age and mass distributions of LMC clusters and conclude that it is hard to constrain the disruption time-scale.  They find that models with a long disruption times ($\gtrsim1\,$Gyr) are needed to describe the mass function, while short disruption times ($\lesssim1\,$Gyr) are preferred for the age distribution. \citet{larsen08}  showed that an exponential truncation in the CIMF around $2\times10^5\,\msun$ provides a good description of the mass function of LMC clusters.
  Perhaps the disruption time-scale of LMC clusters can be better  constrained when the evolved Schechter function from Section~\ref{subsec:nonlin}  is used. 
  
Most probably the model presented in this paper will not be able to explain the age distribution of clusters in the Antennae galaxies \citep{2005ApJ...631L.133F}. The $\tau^{-1}$ scaling of the age distribution and the mass-independent nature of the disruption model invoked to explain this is very peculiar and this has not been found in other galaxies thus far. Since the galactic environment in the Antennae galaxies is quite different than that of a quiescent spiral galaxy, it could be that clusters suffer from different (additional) mechanisms that disrupt clusters of different masses equally fast. Something along these lines was recently suggested by \citet{2008MNRAS.391L..98R} who studied Antennae like mergers with $N$-body simulations. They show that due to the interaction of the two spiral discs there are strong compressive tides that probably induce star and cluster formation. Their models show that {\it if} clusters or associations indeed form in such compressive tides, they can be held together for 10\,Myr or longer, thereby postponing the dissolution of clusters due to gas removal, which in other galaxies seems to destroy the majority ($\sim$90\%) of the embedded clusters within a few Myrs after formation. Although this is speculative, it is important to keep in mind that the cluster system of the Antennae galaxies has very different properties as compared to other cluster systems (see the discussion in \citealt{2008A&A...482..165G}). \citet{larsen08} showed that the CIMF of (normal quiescent) spiral galaxies can be well described by a Schechter function  with $\mc\simeq2\times10^5\,\msun$.  It should be possible to describe most of the age and mass distributions in (quiescent) galaxies with the model presented in this study.

To validate the correctness of the model presented in this study, it would be convincing to `detect' the turnover in the mass function at ages between $\sim$500\,Myr and $\sim$1\,Gyr.
From Fig.~\ref{fig:slopem51} it can be seen that the depth of  observations available at the moment, would have to be increased by roughly one dex in mass, or 2.5 magnitude, to bring the turnover mass, $\mto$, above the detection limit. The improved sensitivity of the new HST/WFC3 camera will allow us to trace the cluster mass function at old ages to the required depth to derive a mass function at intermediate ages down to the turnover ($\mto\simeq10^4\,\msun$) for cluster populations at distances of $5-10$\,Mpc.

\section*{Acknowledgement}
MG enjoyed stimulating discussions with Andr{\'e}s Jord{\'a}n and thanks Nate Bastian, Iraklis Konstantopoulos, Henny Lamers and S{\o}ren Larsen for discussions and  critical reading of earlier versions of this manuscript.
\bibliographystyle{mn2e}

\end{document}